\documentclass[twocolumn,journal]{IEEEtran}
\usepackage[T1]{fontenc}
\usepackage[latin9]{inputenc}
\usepackage{color}
\usepackage{array}
\usepackage{bm}
\usepackage{multirow}
\usepackage{amsmath}
\usepackage{amsthm}
\usepackage{amssymb}
\usepackage{graphicx}
\usepackage[unicode=true,
 bookmarks=true,bookmarksnumbered=true,bookmarksopen=true,bookmarksopenlevel=1,
 breaklinks=false,pdfborder={0 0 0},pdfborderstyle={},backref=false,colorlinks=false]
 {hyperref}
\hypersetup{pdftitle={Your Title},
 pdfauthor={Your Name},
 pdfpagelayout=OneColumn, pdfnewwindow=true, pdfstartview=XYZ, plainpages=false}

\makeatletter

\providecommand{\tabularnewline}{\\}

\theoremstyle{plain}
\newtheorem{thm}{\protect\theoremname}
\theoremstyle{plain}
\newtheorem{prop}[thm]{\protect\propositionname}
\theoremstyle{definition}
\newtheorem{defn}[thm]{\protect\definitionname}
\theoremstyle{plain}
\newtheorem{lem}[thm]{\protect\lemmaname}
\theoremstyle{plain}
\newtheorem{cor}[thm]{\protect\corollaryname}

\usepackage[caption=false,font=footnotesize]{subfig}
\usepackage{diagbox}
\usepackage{algorithm}
\usepackage{algorithmic}
\usepackage{color}
\usepackage{hyperref}
\usepackage{threeparttable}
\newcolumntype{M}{>{$\displaystyle}c<{$}} 
\usepackage{blindtext}

\@ifundefined{showcaptionsetup}{}{%
 \PassOptionsToPackage{caption=false}{subfig}}
\usepackage{subfig}
\makeatother

\providecommand{\corollaryname}{Corollary}
\providecommand{\definitionname}{Definition}
\providecommand{\lemmaname}{Lemma}
\providecommand{\propositionname}{Proposition}
\providecommand{\theoremname}{Theorem}

\begin{document}
\title{An Adaptive and Robust Deep Learning Framework for THz Ultra-Massive MIMO Channel Estimation}
\author{Wentao~Yu,~\IEEEmembership{Graduate Student Member,~IEEE,} Yifei~Shen,~\IEEEmembership{Graduate Student Member,~IEEE,}
Hengtao~He,~\IEEEmembership{Member,~IEEE,} Xianghao~Yu,~\IEEEmembership{Member,~IEEE,}
Shenghui Song,~\IEEEmembership{Senior Member,~IEEE,} Jun~Zhang,~\IEEEmembership{Fellow,~IEEE,}
and~Khaled~B. Letaief,~\IEEEmembership{Fellow,~IEEE}\thanks{An earlier version of this paper was presented in part at the IEEE
Global Communications Conference, Rio de Janeiro, Brazil, Dec. 2022
\cite{2022Yu}. This work was supported in part by the Hong Kong Research
Grants Council under Grant 16212922, 15207220, and the Areas of Excellence
Scheme Grant AoE/E-601/22-R, and in part by a grant from the NSFC/RGC
Joint Research Scheme sponsored by the Research Grants Council of
the Hong Kong Special Administrative Region, China and National Natural
Science Foundation of China (Project No. N\_HKUST656/22). The source
code is publicly available at \protect\href{https://github.com/wyuaq}{https://github.com/wyuaq}.
\textit{(Corresponding author: Jun Zhang.)}}\thanks{Wentao Yu, Hengtao He, Shenghui Song, Jun Zhang, and Khaled B. Letaief
are with the Department of Electronic and Computer Engineering, The
Hong Kong University of Science and Technology, Hong Kong (e-mail:
wyuaq@connect.ust.hk; eehthe@ust.hk; eeshsong@ust.hk; eejzhang@ust.hk;
eekhaled@ust.hk). }\thanks{Yifei Shen is with the Department of Electronic and Computer Engineering,
The Hong Kong University of Science and Technology, Hong Kong, and
also with Microsoft Research Asia, Shanghai 200232, China. (e-mail:
yshenaw@connect.ust.hk). }\thanks{Xianghao Yu is with the Department of Electrical Engineering, City
University of Hong Kong (CityU), Hong Kong (e-mail: alex.yu@cityu.edu.hk). }}
\maketitle
\begin{abstract}
Terahertz ultra-massive MIMO (THz UM-MIMO) is envisioned as one of
the key enablers of 6G wireless networks, for which channel estimation
is highly challenging. Traditional analytical estimation methods are
no longer effective, as the enlarged array aperture and the small
wavelength result in a mixture of far-field and near-field paths,
constituting a hybrid-field channel. Deep learning (DL)-based methods,
despite the competitive performance, generally lack theoretical guarantees
and scale poorly with the size of the array. In this paper, we propose
a general DL framework for THz UM-MIMO channel estimation, which leverages
existing iterative channel estimators and is with provable guarantees.
Each iteration is implemented by a fixed point network (FPN), consisting
of a closed-form linear estimator and a DL-based non-linear estimator.
The proposed method perfectly matches the THz UM-MIMO channel estimation
due to several unique advantages. First, the complexity is low and
adaptive. It enjoys provable linear convergence with a low per-iteration
co\textcolor{black}{st and monotonically increasing accuracy, which
enables an adaptive accuracy-complexity tradeoff. Second, it is robust
to practical distribution shifts and can directly generalize to a
variety of heavily out-of-distribution scenarios with almost no performance
loss, which is suitable for the complicated THz channel conditions.
For practical usage, the proposed framework is further extended to
wideband THz UM-MIMO systems with beam squint effect. Theoretical
analysis and extensive simulation results are provided to illustrate
the advantages over the state-of-the-art methods in estimation accuracy,
convergence rate, complexity, and robustness. }
\end{abstract}


\IEEEpeerreviewmaketitle{}

\section{Introduction}

\IEEEPARstart{T}{he} terahertz (THz) band, ranging from 0.1 to 10
THz, is envisioned as a prime candidate for new spectrum in the sixth
generation (6G) wireless networks \cite{2019Rappaport,2021Sarieddeen}.
It promises to support the explosive demand on wireless traffic \cite{2022Akyildiz},
and lay the foundation for many emerging applications of 6G, such
as extended reality \cite{2022Chaccour} and edge intelligence \cite{2022Letaief}.
Nevertheless, to fully unleash the potential of THz bands, the severe
spread and molecular absorption loss must be compensated to enhance
the coverage range. Thanks to the short wavelength, ultra-massive
multiple-input multiple-output (UM-MIMO) systems can be implemented
with thousands of antennas packed within a small footprint \cite{2022Akyildiz}.
With THz UM-MIMO, highly-directional transmission can be achieved
with advanced beamforming techniques \cite{2020Zhang,2021Ning}, whose
design requires accurate estimation of the high-dimensional channel
with low pilot overhead. 

As a cost-efficient candidate, the planar array-of-subarray (AoSA)
is celebrated as the most promising architecture of THz UM-MIMO, in
which the antenna elements (AEs) are assembled into multiple disjoint
planar subarrays (SAs) \cite{2021Sarieddeen,2016Lin,2018Huang}. The
AEs in each SA share a single radio frequency (RF) chain through dedicated
phase shifters, whose spacings are tiny due to the small wavelength.
By contrast, the SAs are separated by a much larger distance since
integrating too many AEs compactly can reduce the spatial multiplexing
gain and cause difficulties in circuit control and cooling \cite{2021Sarieddeen,2018Huang}.
The advantages of the planar AoSA include a lower hardware cost and
power consumption. In addition, grouping the AEs into SAs increases
the array gain, while the collaboration between SAs provides high
spatial multiplexing gain \cite{2021Sarieddeen,2018Huang}. 

Nevertheless, the planar AoSA architecture also poses several severe
challenges for low-overhead channel estimation. First, due to the
limited RF chains, the received pilot signals are highly compressed
compared with the dimension of the channel, which makes the problem
under-determined. Second, the antenna array is non-uniform owing to
the widely-spaced SAs, while most previous works only considered uniform
arrays. It is difficult to design a unified algorithm that works for
arbitrary array geometry. Most importantly, the enlarged array aperture
and short wavelength of the THz planar AoSAs necessitates near-field
considerations. In general, a dynamic mixture of the far- and near-field
paths coexist and constitute a \textit{hybrid-field} channel \cite{2022Yu,2021Tarboush}.
Nevertheless, existing algorithms mostly assume a uniform array operating
in either the far-field or the near-field region, and thus cannot
satisfactorily address the complicated channel conditions in a practical
THz UM-MIMO system with planar AoSA. 

\subsection{Related Works\label{subsec:Related-Works}}

Due to the limited number of RF chains, the pilot overhead of traditional
least squares (LS) estimators is very high since they require the
received pilots to have at least the same dimension as the channel
to ensure robust estimation. Prior knowledge of the channel should
be exploited in order to reduce the pilot overhead. Many compressed
sensing (CS)-based algorithms were investigated to serve the purpose,
which can be grouped into three categories, i.e., sparse reconstruction-based,
Bayesian, and deep learning (DL)-based methods. 

The key of sparse reconstruction-based methods is to design a proper
dictionary matrix to represent the channel as a sparse vector. Once
the dictionary is determined, channel estimation can be transformed
into a sparse reconstruction problem and many off-the-shelf algorithms
are readily applicable \cite{2017Choi}. For a uniform array operating
in the far field, the channel is sparse under a discrete Fourier transform
(DFT)-based dictionary \cite{2021Dovelos}, which corresponds to sampling
the angular domain by uniform grids. In the near field, the array
response is affected by both angle and distance, which requires sampling
both domains via dedicated grid patterns \cite{2022Cui}. Previous
works revealed that the far-field dictionary cannot properly sparsify
the near-field channel, and vice versa \cite{2022Wei}. The optimal
hybrid-field dictionary is thus dependent on the portion of path components,
whose design is still an open problem. For a non-uniform array operating
in the hybrid-field mode, e.g., the considered problem, the only option
in the literature is dictionary learning \cite{2018Ding}, which optimizes
the dictionary using a large \textit{site-specific} dataset, but the
learned dictionary can generalize poorly on other sites. 

Bayesian methods instead depend on the prior distribution of the channel.
If the prior is available, Bayesian-optimal estimation can be achieved
by iterative algorithms with affordable complexity, e.g., the approximate
message passing (AMP) and its variants \cite{2009Donoho,2017Ma}.
However, since the true prior distribution is unknown in practice,
previous works first empirically chose a base distribution and then
updated the distribution parameters during each iteration via the
expectation-maximization (EM) principle. The chosen prior distributions
can be either unstructured, such as Gaussian mixture \cite{2015Wen,2022Srivastava},
Laplacian \cite{2019Bellili,2019Wang}, or Bernoulli-Gaussian distributions
\cite{2019Huang}, or structured if fine-grained knowledge of the
channel is available, such as hidden Markov model \cite{2018Liu}.
As a rule of thumb, \textit{matched} and structured priors often offer
higher estimation accuracy than unstructured ones. Nevertheless, the
former depends on delicate channel models with stronger assumptions,
which may not be available for the complicated hybrid-field THz UM-MIMO
channel. The latter can be adopted for arbitrary channel conditions,
but may suffer from an inferior accuracy. 

DL-based methods, on the other hand, do not depend on the structural
information of the channel, but rather learn to \textcolor{black}{exploit
it from data, which makes them suitable to handle the complicated
channel conditions in which analytical methods perform poorly. Existing
methods can be categorized as data-driven and model-driven ones \cite{2019He}.
The former learns a direct mapping from the received pilots to the
estimated full channel or its parameters by using a pure neural network
model \cite{2019Dong,2021ChenTCOMM}. Although being time-efficient,
the performance is often inferior since the information contained
in the measurement matrix cannot be effectively utilized \cite{2018He}.
By contrast, model-driven DL-based methods, also known as deep unfolding,
are built by }\textit{\textcolor{black}{truncatin}}\textcolor{black}{g
an iterative algorithm into }\textit{\textcolor{black}{finite}}\textcolor{black}{{}
and }\textit{\textcolor{black}{fixed}}\textcolor{black}{{} layers and
replace the bottleneck modules in each layer with learnable components,
which can offer better performance and interpretability compared with
data-driven methods \cite{2018He,2022He,2022Hu,2021Jin}. Besides
channel estimation, deep unfolding has also found successful applications
in other physical layer problems, such as data detection and decoding
\cite{2019Alexios}. }

Despite their superior performance, existing deep unfolding methods
suffer from multiple crucial drawbacks, which makes them unsuitable
for THz UM-MIMO. First, the \textbf{scalability }is poor. Training
unfolded algorithms entails tracking the intermediate states and gradients
per layer, which leads to a huge memory and computational overhead,
and lacks scalability with large-scale array and deeper layers. Second,
the \textbf{complexity} is high and not adaptive. Different from the
classical algorithms which can support an \textit{adaptive} number
of iterations, unfolded algorithms emit the estimation after a \textit{pre-defined}
and \textit{fixed} number of layers. Third, the \textbf{reliability
}is not guaranteed. Although unfolded algorithms are meant to emulate
classical algorithms which iteratively refine the estimation and output
the \textit{fixed point} after an adaptive number of iterations, they
stop the after the \textit{pre-defined} number of layers \textit{without}
theoretical guarantees. The unfolded algorithm is often oscillating
rather than converging \cite{2022Chen-tianlong}. Most importantly,
the \textbf{generalization }issue is not well addressed. A critical
drawback of DL-based methods is the risk of performance degradation
in \textit{out-of-distribution} scenarios, where the distributions
of the channel, measurement matrices, and noise in the testing environment
may differ from those seen during training. This can frequently happen
in THz UM-MIMO systems due to, e.g., the dynamic portion of hybrid-field
paths and line-of-sight (LoS) blockage. However, these important issues
are largely overlooked in the literature, which motivates our work. 

\subsection{Contributions}

In this paper, we propose a \textit{unified} and \textit{theoretically
sound} DL framework called fixed point networks (FPNs) to tackle all
the critical drawbacks mentioned above, achieving a low-complexity,
adaptive, and robust estimation of the highly complicated hybrid-field
THz UM-MIMO channel with a general non-uniform planar AoSA. Our contributions
are as follows. 
\begin{itemize}
\item To tackle the drawbacks of existing DL-based algorithms, we propose
FPNs as a general framework that constructs channel estimation algorithms
as the fixed point iteration of a contraction mapping, comprising
a closed-form linear estimator from classical algorithms and a DL-based
non-linear estimator. The estimated channel is the fixed point of
the contraction, which can be efficiently calculated. 
\begin{figure*}[t]
\centering{}\subfloat[\label{fig:Array-of-subarray}]{
\centering{}\includegraphics[width=0.23\textwidth]{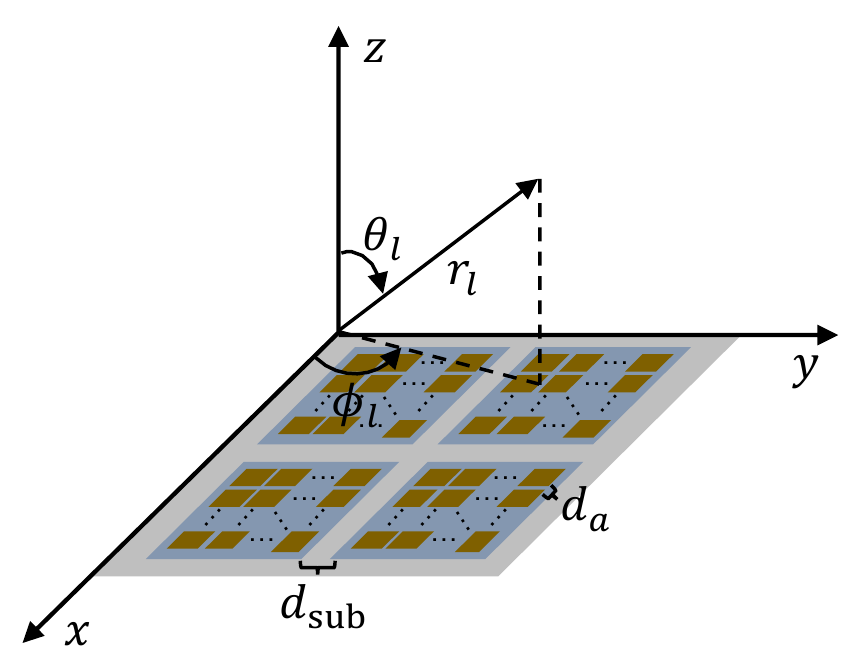}}\quad \quad\subfloat[\label{fig:Partially-connected}]{\centering{}\includegraphics[width=0.23\textwidth]{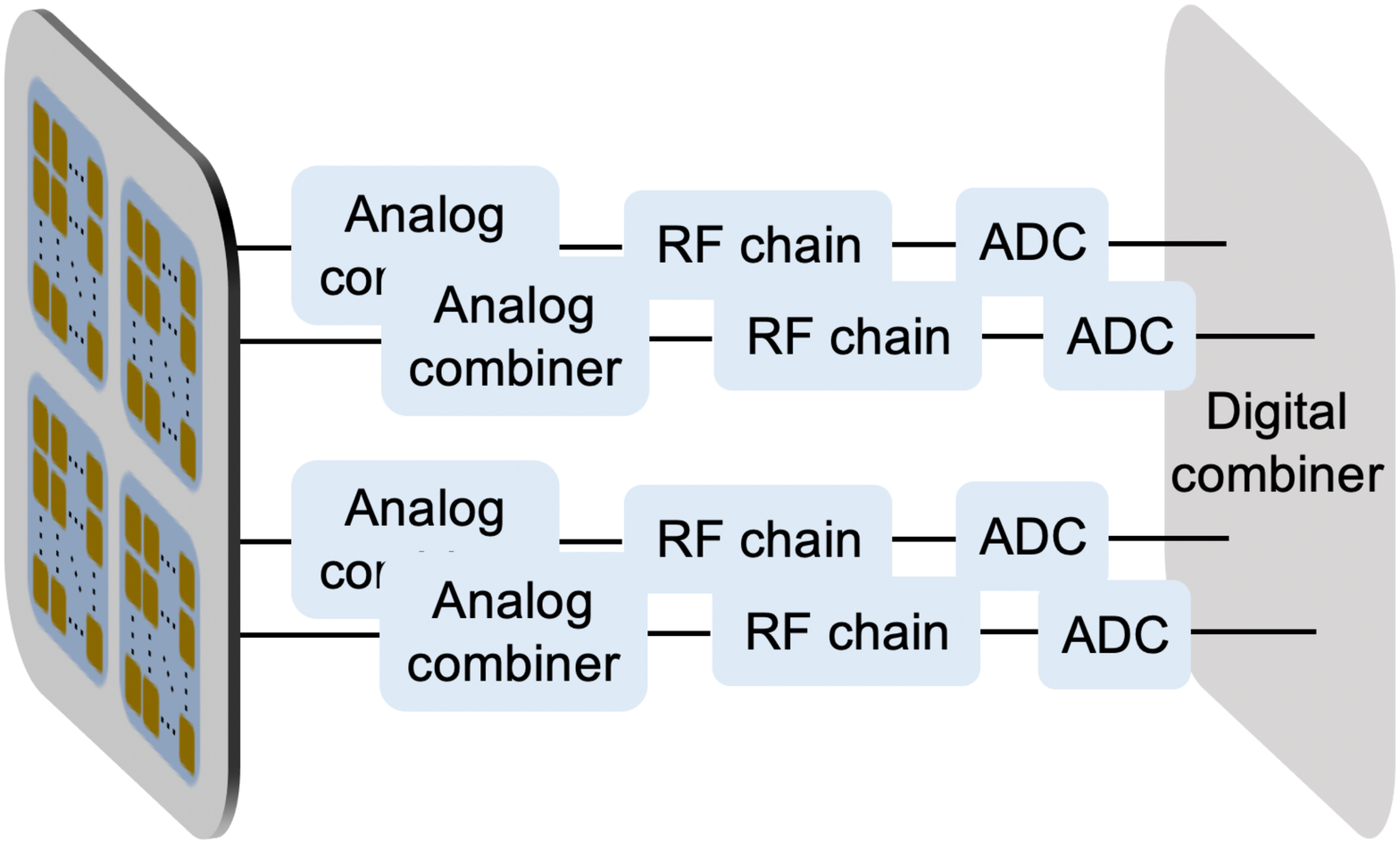}}\quad \quad\subfloat[\label{fig:Hybrid-field}]{\centering{}\includegraphics[width=0.23\textwidth]{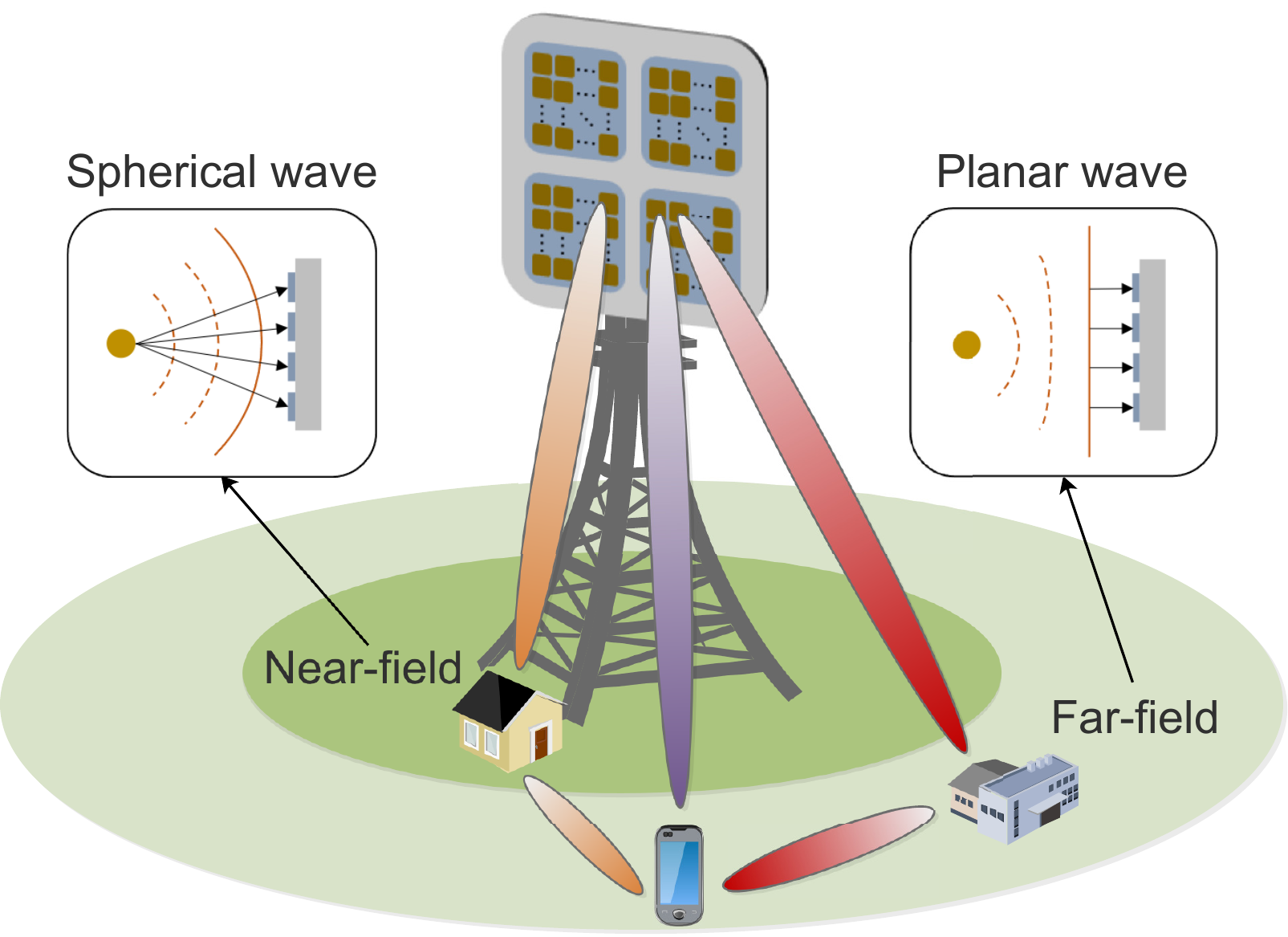}}\caption{System model. (a) Planar AoSA geometry of the THz UM-MIMO, in which
SAs are denoted by dark blue squares while AEs are denoted by dark
golden squares. (b) Partially-connected hybrid analog-digital combining
in the AoSA. The AEs in each SA share the same RF chain through dedicated
phase shifters. (c) A typical hybrid far- and near-field propagation
environment. The wavefront is spherical in the near field, while is
planar in the far field. \label{fig:System-model}}
\end{figure*}
\item We show the compatibility of FPNs with a wide range of iterative algorithms
along with general design guidelines, and propose FPN-OAMP, an FPN-enh\textcolor{black}{anced
estimator via orthogonal approximate message passing (OAMP) \cite{2017Ma}.
The proposed algorithm is applicable to hybrid-field THz UM-MIMO systems
operating in not only the narrowband mode but also the wideband one
with beam squint \cite{2018WangTSP}. }
\item The theoretical benefits of our proposed method are two-fold. First,
it scales gracefully with the size of UM-MIMO since the training via
one-step gradient requires only \textit{constant} complexity regardless
of the specific number of iterations \cite{2022Fung}. Second, thanks
to the nice properties of contraction, it can run an arbitrary number
of iterations with a provable linear convergence rate and monotonically
increasing accuracy. This not only ensures the reliability but also
offers an \textit{adaptive} accuracy-complexity tradeoff. Simulation
results support the theoretical properties, and show the \textit{fastest}
convergence rate and the \textit{state-of-the-art} performance compared
to existing iterative estimators. 
\item Our proposed method also enjoys the empirical benefit of out-of-distribution
\textit{robustness}. With extensive simulation results, we confirm
that it can enjoy \textit{direct} generalization to heavy distribution
shifts in the channel, measurement matrix, and noise with only negligible
performance loss. Notably, our method can also generalize to different
array geometries. For the rare cases where direct generalization fails,
we propose an unsupervised \textit{self-adaptation} scheme to enable
online adaptation to abrupt shifts. 
\end{itemize}

\subsection{Paper Organization and Notation}

The remaining parts of the paper are organized as follows. In Section
\ref{sec:System-Model-and}, we introduce the system model, the channel
model, and the problem formulation. In Section \ref{sec:Fixed-Point-Networks:},
we explain the motivation and the general ideas behind the proposed
FPNs, and design a specific FPN-enhanced channel estimator based on
OAMP, i.e., FPN-OAMP, and prove the key theoretical properties. In
Sections \ref{sec:Simulation-Results} and \ref{sec:Robust-Out-of-Distribution-Perfo},
extensive simulation results are provided to illustrate the advantages
of our proposal in both the performance and the out-of-distribution
robustness. 

\textit{Notation:} $|a|$ is the absolute value of scalar $a$. $\|\mathbf{a}\|_{p}$
and $(\mathbf{a})_{i}$ are the $\ell_{p}$-norm and the $i$-th element
of vector $\mathbf{a}$, respectively. $\mathbf{A}^{T}$, $\mathbf{A}^{H}$,
$\mathbf{A^{\dagger}}$, $\text{tr}(\mathbf{A})$, $\text{vec}(\mathbf{A})$,
$\text{\ensuremath{\Re}}(\mathbf{A})$, $\ensuremath{\Im}(\mathbf{A})$,
$(\mathbf{A})_{i,j}$ are the transpose, Hermitian, pseudo-inverse,
trace, vectorization, real part, imaginary part, and the $(i,j)$-th
element of matrix $\mathbf{A}$, respectively. $\mathbf{B}=\text{blkdiag}(\mathbf{A}_{1},\mathbf{A}_{2},\ldots,\mathbf{A}_{n})$
is a block diagonal matrix by aligning $\mathbf{A}_{1},\mathbf{A}_{2},\ldots,\mathbf{A}_{n}$
on the diagonal. $\circ$ denotes the composition of functions. $\mathcal{U}(a_{1},a_{2})$
is a uniform distribution over the interval $[a_{1},a_{2}]$. $\mathcal{CN}(\boldsymbol{\mu},\mathbf{R})$
is a complex Gaussian distribution with mean $\boldsymbol{\mu}$ and
covariance $\mathbf{R}$. 

\section{System Model and Problem Formulation\label{sec:System-Model-and}}

\subsection{System Model}

We consider the uplink channel estimation problem for THz UM-MIMO
systems. The base station (BS) is equipped with a planar AoSA with
$\sqrt{S}\times\sqrt{S}$ SAs. Each SA is a uniform planar array (UPA)
consisting of $\sqrt{\bar{S}}\times\sqrt{\bar{S}}$ AEs, as illustrated
in Fig. \ref{fig:System-model}(a). The BS has a total of $S\bar{S}$
antennas. To improve the energy efficiency, the AoSA adopts the partially-connected
hybrid beamforming \cite{2020Zhang,2016Lin}, as shown in Fig. \ref{fig:System-model}(b).
Within each SA, the AEs share the same RF chain via dedicated phase
shifters. A total of $S$ RF chains are utilized to receive data streams
from multiple single-antenna user equipments (UEs)\textcolor{black}{}\footnote{\textcolor{black}{The system model for multi-antenna UEs is similar
to the single-antenna case after some proper vectorization. Please
see Section \uppercase\expandafter{\romannumeral5} in \cite{2021Dovelos}
for details. However, if the number of antennas at the UEs is comparable
to that at the BS, the dual-side near-field effect needs to be considered
\cite{2023Cui}, which will further complicate the hybrid-field channel
model. Nevertheless, it is still possible to extend the proposed framework
to such a case since the system models are similar, while the channel
distributions can be learned from data. We leave it as a future direction
due to the limited space. Although we have focused on the uplink channel
estimation to illustrate the algorithm, extension to the downlink
channel estimation can be performed in a similar manner as \cite{2021Ma}. }}. 

We define the index $s$ of the SA at the $m$-th row and $n$-th
column of the AoSA by $s=(m-1)\sqrt{S}+n$, where $1\leq m,n\leq\sqrt{S}$
and $1\leq s\leq S$. Similarly, we define the index $\bar{s}$ of
the AE at the $\bar{m}$-th row and $\bar{n}$-th column of a certain
SA by $\bar{s}=(\bar{m}-1)\sqrt{\bar{S}}+\bar{n}$, where $1\leq\bar{m},\bar{n}\leq\sqrt{\bar{S}}$
and $1\leq\bar{s}\leq\bar{S}$. The distances between adjacent SAs
and adjacent AEs are denoted by $d_{\text{sub}}$ and $d_{a}$, respectively.
As shown in Fig. \ref{fig:System-model}(a), we construct a Cartesian
coordinate system with the origin point being the first AE in the
first SA. Assuming that the AoSA lies in the $x$-$y$ plane, then
the coordinate of the $\bar{s}$-th AE in the $s$-th SA is given
by
\begin{equation}
\mathbf{p}_{s,\bar{s}}=\left(\begin{array}{c}
(m-1)[(\sqrt{\bar{S}}-1)d_{a}+d_{\text{sub}}]+(\bar{m}-1)d_{a}\\
(n-1)[(\sqrt{\bar{S}}-1)d_{a}+d_{\text{sub}}]+(\bar{n}-1)d_{a}\\
0
\end{array}\right).
\end{equation}
The array aperture of the planar AoSA, denoted by $D$, equals the
length of its diagonal, and is given by
\begin{equation}
\begin{alignedat}{1}D & =\|\mathbf{p}_{1,1}-\mathbf{p}_{S,\bar{S}}\|_{2}\\
 & =\sqrt{2}[\sqrt{S}(\sqrt{\bar{S}}-1)d_{a}+(\sqrt{S}-1)d_{\text{sub}}],
\end{alignedat}
\label{eq:array-aperture}
\end{equation}
\textcolor{black}{where the distance between adjacent AEs is configured
as half the carrier wavelength $\lambda_{c}$, i.e., $d_{a}=\frac{\lambda_{c}}{2}$,
while the distance between adjacent SAs is given by $d_{\text{sub}}=wd_{a}\,(w\gg1)$,
since we mainly consider widely spaced SAs that are typical in THz
UM-MIMO systems \cite{2018Huang,2021Tarboush,2021ChenTCOMM}. }

\textcolor{black}{We then introduce the far-field and near-field considerations
in THz UM-MIMO systems. According to the distance from the RF source
to the antenna array, wave propagation can be divided into the far-field
and the near-field regions \cite{2021Tarboush,2021Dovelos}. As shown
in Fig. \ref{fig:System-model}(c), in the far-field region, the wavefront
is approximately planar and the angle of arrival (AoA) at each AE
can be assumed equal. In the near field, however, the planar wave
assumption no longer holds. In such a case, the spherical wavefront
must be considered in channel modeling. }

\textcolor{black}{The boundary between the far- and near-field regions
is the Rayleigh (or Fraunhofer) distance, i.e., $D_{\text{Rayleigh}}=\frac{2D^{2}}{\lambda_{c}}$
\cite{2015Balanis}, which is related to both the carrier wavelength
and the array geometry. Plugging (\ref{eq:array-aperture}) into the
expression, we obtain that 
\begin{equation}
\begin{alignedat}{1}D_{\text{Rayleigh}} & =\{\sqrt{S}(\sqrt{\bar{S}}-1)+(\sqrt{S}-1)w\}^{2}\lambda_{c}\end{alignedat}
.
\end{equation}
Although $\lambda_{c}$ is small in THz systems, the massive number
of AEs and the widely spaced SAs can still result in a quite large
near-field region. Due to the limited coverage of THz wave, the far-field
and near-field regions typically coexist. The portion can vary based
on the specific system settings. For reference, $D_{\text{Rayleigh}}$
in some typical THz UM-MIMO systems with planar AoSA is illustrated
in Fig. \ref{fig:Accuracy-of-the}. }

In view of this, we study the general case where the channel consists
of a mixture of far- and near-field paths, i.e., the \textit{hybrid-field}
condition. In addition, the considered non-uniform planar AoSA also
represents a general class of array geometry for THz UM-MIMO systems.
The method proposed in this work is thus applicable to a broad range
of system settings. 

\subsection{Hybrid-Field THz UM-MIMO Channel Model \label{subsec:Hybrid-Field-THz-UM-MIMO}}

Due to the limited scattering of the THz wave, the spati\textcolor{black}{al-frequency
channel $\mathbf{\tilde{h}}\in\mathbb{C}^{S\bar{S}\times1}$ between
the BS and a specific UE can be characterized by the superposition
of one LoS path and $L-1$ non-LoS paths \cite{2021Dovelos}, given
by
\begin{equation}
\mathbf{\tilde{h}}=\sum_{l=1}^{L}\alpha_{l}(f_{c})\mathbf{a}\left(\phi_{l},\theta_{l},r_{l},f_{c}\right)e^{-j2\pi f_{c}\tau_{l}}\label{eq:channel}
\end{equation}
where $f_{c}$ denotes the carrier frequency, while $\alpha_{l}(f_{c})$,
$\phi_{l}$, $\theta_{l}$, $r_{l}$, $\mathbf{a}\left(\phi_{l},\theta_{l},r_{l},f_{c}\right)$,
and $\tau_{l}$ are respectively the path loss, azimuth AoA, elevation
AoA, distance between the array and the RF source/scatterer, the array
response vector, and the time delay of the $l$-th path. In particular,
$\phi_{l}$, $\theta_{l}$, and $r_{l}$ are measured with respect
to the origin of the coordinate system, as shown in Fig. \ref{fig:System-model}(a). }

\subsubsection{Path Loss}

In addition to the spread loss, the molecular absorption loss is non-negligible
at the THz band. The path loss $\alpha_{l}$ accounts for both of
them. Assuming that $l=1$ denotes the LoS path and $l>1$ denotes
the NLoS paths, then
\begin{equation}
\alpha_{l}(f_{c})=|\Gamma_{l}|\left(\frac{c}{4\pi f_{c}r_{1}}\right)e^{-\frac{1}{2}k_{\text{abs}}(f_{c})r_{1}},
\end{equation}
where $\Gamma_{l}$ is the reflection coefficient, $r_{1}$ is the
LoS path length, and $k_{\text{abs}}$ is the molecular absorption
coefficient \cite{2021Dovelos}. For the LoS path, $\Gamma_{l}=1$.
For NLoS paths, $\Gamma_{l}$ is given by
\begin{equation}
\Gamma_{l}=\frac{\cos\varphi_{\text{in},l}-n_{t}\cos\varphi_{\text{ref},l}}{\cos\varphi_{\text{in},l}+n_{t}\cos\varphi_{\text{ref},l}}e^{-\left(\frac{8\pi^{2}f_{c}^{2}\sigma_{\text{rough}}^{2}\text{cos}^{2}\varphi_{\text{in},l}}{c^{2}}\right)},
\end{equation}
where $\varphi_{\text{in},l}$ is the angle of incidence of the $l$-th
path, $\varphi_{\text{ref},l}=\arcsin(n_{t}^{-1}\sin\varphi_{\text{in},l})$
is the angle of refraction. Also, $n_{t}$ and $\sigma_{\text{rough}}$
are respectively the refractive index and the roughness coefficient
of the reflecting material \cite{2021Dovelos}. Due to the severe
penetration of the diffused and diffracted rays at THz band, their
contributions are negligible over only a few meters \cite{2022Akyildiz}.
Therefore, similar to \cite{2021Dovelos,2021Tarboush}, the NLoS path
loss model takes into account the single-bounce reflected rays only. 

\subsubsection{Array Response Vector}

As mentioned before, the array response vector $\mathbf{a}^{\text{}}(\phi_{l},\theta_{l},r_{l},f_{c})\in\mathbb{C}^{S\bar{S}\times1}$
differs in the far- and near-field regions, which are determined by
the Rayleigh distance $D_{\text{Rayleigh}}$, and is given by 
\begin{equation}
\mathbf{a}(\phi_{l},\theta_{l},r_{l},f_{c})=\begin{cases}
\mathbf{a}^{\text{near}}(\phi_{l},\theta_{l},r_{l},f_{c}), & \text{if }r_{l}<D_{\text{Rayleigh}},\\
\mathbf{a}^{\text{far}}(\phi_{l},\theta_{l},r_{l},f_{c}), & \text{otherwise}.
\end{cases}
\end{equation}
For notational brevity, we first construct the array response matrix.
Due to the spherical wavefront, each element of the near-field array
response matrix depends on the \textit{exact} distance between each
AE and the RF source/scatterer. For the $l$-th path, the position
of the RF source/scatterer is $r_{l}\mathbf{t}_{l}$, where $\mathbf{t}_{l}$
is the unit-length vector in the AoA direction, given by $\mathbf{t}_{l}=(\sin\theta_{l}\cos\phi_{l},\sin\theta_{l}\sin\phi_{l},\cos\theta_{l})^{T}$.
Therefore, the array response of the $\bar{s}$-th AE in the $s$-th
SA is given by
\begin{equation}
(\mathbf{A}^{\text{near}}(\phi_{l},\theta_{l},r_{l},f_{c}))_{s,\bar{s}}=e^{-j2\pi\frac{f_{c}}{c}\|\mathbf{p}_{s,\bar{s}}-r_{l}\mathbf{t}_{l}\|_{2}},\label{eq:near-field-response}
\end{equation}
where $c$ is the speed of light. The near-field array response vector
is $\mathbf{a}^{\text{near}}(\phi_{l},\theta_{l},r_{l},f_{c})=\text{vec}(\mathbf{A}^{\text{near}}(\phi_{l},\theta_{l},r_{l},f_{c}))$.
Due to the planar wavefront in the far-field region, the exact distance
can be approximated by a linear function of the SA and AE indexes.
Therefore, each element of the far-field array response matrix $\mathbf{A}^{\text{far}}(\phi_{l},\theta_{l},r_{l},f_{c})$
is given by
\begin{figure}[t]
\centering{}\includegraphics[width=6cm]{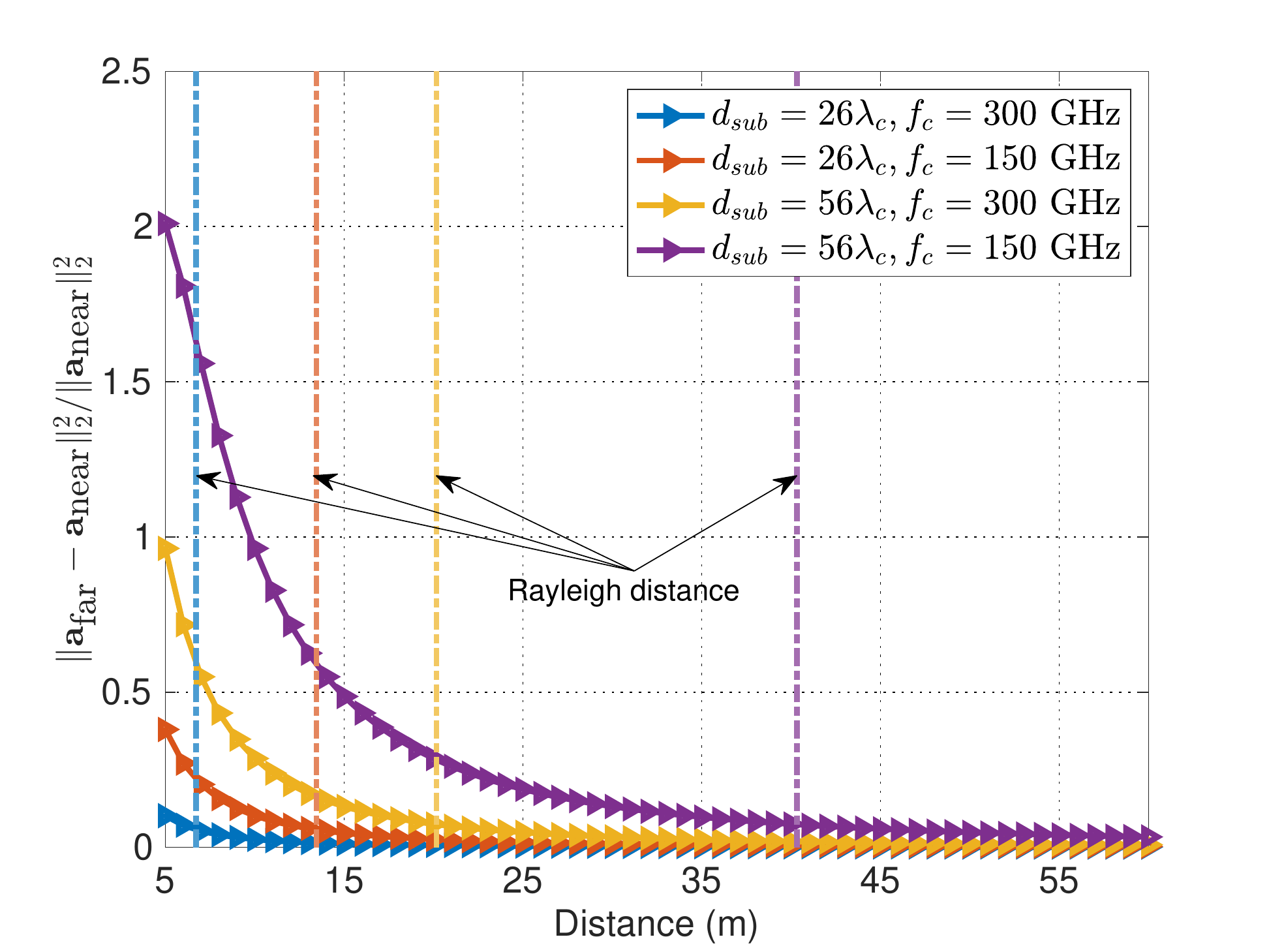}\caption{Accuracy of the far-field array response versus the distance $r_{l}$
using (\ref{eq:near-field-response}) and (\ref{eq:far-field-response}),
when $S=4$, $\bar{S}=256$, $d_{a}=\lambda_{c}/2$, $\theta_{l}=0.4\pi$,
and $\phi_{l}=-0.7\pi$. \label{fig:Accuracy-of-the}}
\end{figure}
\begin{table*}[t]
\begin{centering}
\caption{Per-iteration update rules for representative iterative channel estimators\label{tab:Per-iteration-update-rules}}
\par\end{centering}
\footnotesize
\centering{}%
\begin{tabular}{>{\centering}p{3.5cm}|c|c}
\hline 
\textbf{Algorithm} & \textbf{Linear estimator (LE)} & \textbf{Non-linear estimator (NLE)}\tabularnewline
\hline 
\multirow{1}{3.5cm}{\centering{}PGD} & \multirow{1}{*}{$\mathbf{u}^{(t+1)}=\mathbf{h}^{(t)}+\rho\mathbf{M}^{H}(\mathbf{y}-\mathbf{M}\mathbf{h}^{(t)})$} & $\mathbf{h}^{(t+1)}=\text{Prox}_{1/\rho R}(\mathbf{u}^{(t+1)})$\tabularnewline
\hline 
\multirow{2}{3.5cm}{\centering{}ADMM} & \multirow{2}{*}{$\mathbf{h}^{(t)}=(\mathbf{M}^{H}\mathbf{M}+\rho\mathbf{I})^{-1}(\mathbf{M}^{H}\mathbf{y}+\mathbf{u}^{(t)})$} & $\mathbf{z}^{(t)}=\text{Prox}_{1/\rho R}(2\mathbf{h}^{(t)}-\mathbf{u}^{(t)}/\rho)$\tabularnewline
 &  & $\mathbf{u}^{(t+1)}=\mathbf{u}^{(t)}+2\rho(\mathbf{z}^{(t)}-\mathbf{h}^{(t)})$\tabularnewline
\hline 
OAMP & $\mathbf{u}^{(t+1)}=\mathbf{h}^{(t)}+\mathbf{W}^{(t)}(\mathbf{y}-\mathbf{M}\mathbf{h}^{(t)})$ & $\mathbf{h}^{(t+1)}=\eta_{t}(\mathbf{u}^{(t+1)})$\tabularnewline
\hline 
\end{tabular}
\end{table*}
\begin{equation}
(\mathbf{A}^{\text{far}}(\phi_{l},\theta_{l},r_{l},f_{c}))_{s,\bar{s}}=e^{-j2\pi\frac{f_{c}}{c}(\mathbf{p}_{s,\bar{s}}^{T}\mathbf{t}_{l}-r_{l})}.\label{eq:far-field-response}
\end{equation}
where the term $(\mathbf{p}_{s,\bar{s}}^{T}\mathbf{t}_{l}-r_{l})$
in the exponent is the linear approximation of the exact distance
$\|\mathbf{p}_{s,\bar{s}}-r_{l}\mathbf{t}_{l}\|_{2}$ obtained by
the first-order Taylor expansion. Afterwards, the relevant far-field
array response vector can also be obtained by vectorization, i.e.,
$\mathbf{a}^{\text{far}}(\phi_{l},\theta_{l},r_{l},f_{c})=\text{vec}(\mathbf{A}^{\text{far}}(\phi_{l},\theta_{l},r_{l},f_{c}))$. 

In Fig. \ref{fig:Accuracy-of-the}, we plot the accuracy of the far-field
array response versus the distance $r_{l}$, when $S=4$, $\bar{S}=256$,
$d_{a}=\lambda_{c}/2$, $\theta_{l}=0.4\pi$, and $\phi_{l}=-0.7\pi$.
Four curves with different SA spacings and carrier frequencies are
drawn for comparison. Depending on the system settings, the portion
of the far- and near-field regions vary. The Rayleigh distance is
shown by a vertical line with the same color as the corresponding
curve. We can see that the far-field array response is correct when
$r_{l}$ is beyond the Rayleigh distance $D_{\text{Rayleigh}}$. However,
the approximation is inappropriate in the near-field region where
$r_{l}<D_{\text{Rayleigh}}$. In this case, the accurate near-field
array response must be used. 

\subsubsection{Sparse Channel Representation}

For channel estimators based on sparse reconstruction, the fundamental
assumption is that the spatial channel $\mathbf{\tilde{h}}$ can be
transformed to its sparse representation\footnote{We introduce the sparse representation here only for ease of comparison
with the benchmarks. Our proposal does not rely on the channel sparsity. } $\mathbf{\bar{h}}$ in the form of $\mathbf{\tilde{h}}=\mathbf{F}\mathbf{\bar{h}}$,
under an appropriate dictionary matrix $\mathbf{F}$. Since each SA
in the planar AoSA is a UPA with identical geometry, we could design
the overall dictionary matrix in an SA-by-SA manner. For \textit{far-field}
paths, the array response in (\ref{eq:far-field-response}) is a linear
function of the AE index, and is insensitive to the distance $r_{l}$.
Therefore, for each SA, one can use the DFT-based dictionary $\mathbf{U}$
to uniformly sample the AoAs $\theta_{l}$ and $\phi_{l}$, which
is constructed by the Kronecker product of two normalized DFT matrices
of size $\sqrt{\bar{S}}\times\sqrt{\bar{S}}$ \cite{2018Ding}. The
overall dictionary matrix is then $\mathbf{F}=\text{blkdiag}(\mathbf{U}_{1},\mathbf{U}_{2},\ldots,\mathbf{U}_{S})$
with each component matrix $\mathbf{U}_{s}=\mathbf{U}$. For \textit{near-field}
paths, the array response in (\ref{eq:near-field-response}) is a
non-linear function of the AE index, and is sensitive to both the
AoAs $\theta_{l}$, $\phi_{l}$ and the distance $r_{l}$. To handle
this, the idea in previous works is to sample both the AoAs and the
distance to construct a higher dimensional angle-distance domain dictionary
matrix \cite{2022Cui}. However, for the considered \textit{hybrid-field}
case, the optimal dictionary is dependent on the portion of the far-field
and near-field paths. The state-of-the-art solution in the literature
is to apply dictionary learning to optimize the dictionary matrix
$\mathbf{F}$ using site-specific data \cite{2018Ding}. 

In our simulations, we adopt dictionary learning when comparing with
the CS-based benchmarks since their performance is heavily affected
by the quality of the sparse representation. Discussions on the details
are deferred to Section \ref{sec:Simulation-Results}. For the proposed
method and the other DL-based benchmarks, we simply adopt the DFT-based
far-field dictionary since their performance does not rely on the
channel sparsity. 

\subsection{Problem Formulation}

In the uplink channel estimation, the UEs transmit training pilots
to the BS for $Q$ time slots. We assume that orthogonal pilots are
adopted and consider an arbitrary UE without loss of generality \cite{2021Dovelos,2022Cui}.
The received pilot signal $\mathbf{y}_{q}\in\mathbb{C}^{S\times1}$
in the $q$-th time slot at the BS is given by 
\begin{equation}
\begin{aligned}\mathbf{y}_{q} & =\mathbf{W}_{\text{BB},q}^{H}\mathbf{W}_{\text{RF},q}^{H}(\mathbf{\tilde{h}}s_{q}+\mathbf{n}_{q}),\\
 & =\mathbf{W}_{\text{BB},q}^{H}\mathbf{W}_{\text{RF},q}^{H}\mathbf{F}\mathbf{\bar{h}}s_{q}+\mathbf{W}_{\text{BB},q}^{H}\mathbf{W}_{\text{RF},q}^{H}\mathbf{n}_{q},
\end{aligned}
\end{equation}
where $\mathbf{W}_{\text{BB},q}\in\mathbb{C}^{S\times S}$ denotes
the digital combining matrix, $\mathbf{W}_{\text{RF},q}\mathbf{=\text{blkdiag\ensuremath{\left(\mathbf{w}_{1,q},\mathbf{w}_{2,q},\ldots\mathbf{w}_{S,q}\right)}}\in\mathbb{C}}^{S\bar{S}\times S}$
is the analog combining matrix where the elements of each component
vector $\mathbf{w}_{i,q}\in\mathbb{C}^{\bar{S}\times1}$ satisfy the
constant-modulus constraint, $s_{q}$ is the pilot symbol that is
set as 1 for convenience, and $\mathbf{\mathbf{n}}_{q}\sim\mathcal{CN}(\mathbf{0},\sigma_{n}^{2}\mathbf{I})$
is the additive white Gaussian noise (AWGN). Since the combining matrices
cannot be optimally tuned without knowledge of the channel, we consider
an arbitrary scenario where $\mathbf{W}_{\text{BB},q}$ is set as
identity $\mathbf{I}$ and the analog phase shifts in $\mathbf{W}_{\text{RF},q}$
are randomly picked from \textit{one-bit} quantized angles, i.e.,
$(\mathbf{w}_{i,q})_{j}\in\frac{1}{\sqrt{\bar{S}}}\{\pm1\}$, to reduce
the energy consumption \cite{2018He}. The received pilot signal $\mathbf{\bar{y}}=[\mathbf{y}_{1}^{T},\mathbf{y}_{2}^{T},\ldots,\mathbf{y}_{Q}^{T}]^{T}\in\mathbb{C}^{SQ\times1}$
after $Q$ time slots of the training pilot transmission is given
by $\mathbf{\bar{y}}=\text{\ensuremath{\mathbf{\bar{M}}}}\mathbf{\bar{h}}+\mathbf{\bar{n}}$,
where $\bar{\mathbf{M}}=[(\mathbf{W}_{\text{RF},1}^{H}\mathbf{F})^{T},\ldots,(\mathbf{W}_{\text{RF},Q}^{H}\mathbf{F})^{T}]^{T}\in\mathbb{C}^{SQ\times S\bar{S}}$,
and $\mathbf{\bar{n}}=[(\mathbf{W}_{\text{RF},1}^{H}\mathbf{n}_{1})^{T},\ldots,(\mathbf{W}_{\text{RF},Q}^{H}\mathbf{n}_{Q})^{T}]^{T}\in\mathbb{C}^{SQ\times1}$. 

Since deep learning packages require real-valued inputs, we transform
the system model into its equivalent form. Denoting $\mathbf{y}=[\text{\ensuremath{\Re}}(\mathbf{\bar{y}})^{T},\text{\ensuremath{\Im}}(\mathbf{\bar{y}})^{T}]^{T}\in\mathbb{R}^{2SQ\times1}$,
$\mathbf{h}=[\text{\ensuremath{\Re}}(\mathbf{\bar{h}})^{T},\text{\ensuremath{\Im}}(\mathbf{\bar{h}})^{T}]^{T}\in\mathbb{R}^{2S\bar{S}\times1}$.
$\mathbf{n}=[\text{\ensuremath{\Re}}(\mathbf{\bar{n}})^{T},\text{\ensuremath{\Im}}(\mathbf{\bar{n}})^{T}]^{T}\in\mathbb{R}^{2SQ\times1}$,
and
\begin{equation}
\mathbf{M}=\left(\begin{array}{cc}
\text{\ensuremath{\Re}}(\mathbf{\bar{M}}) & -\text{\ensuremath{\Im}}(\mathbf{\bar{M}})\\
\text{\ensuremath{\Im}}(\mathbf{\bar{M}}) & \text{\ensuremath{\Re}}(\mathbf{\bar{M}})
\end{array}\right)\in\mathbb{R}^{2SQ\times2S\bar{S}},
\end{equation}
the equivalent real-valued system model is given by
\begin{figure*}[t]
\centering{}\includegraphics[width=0.75\textwidth]{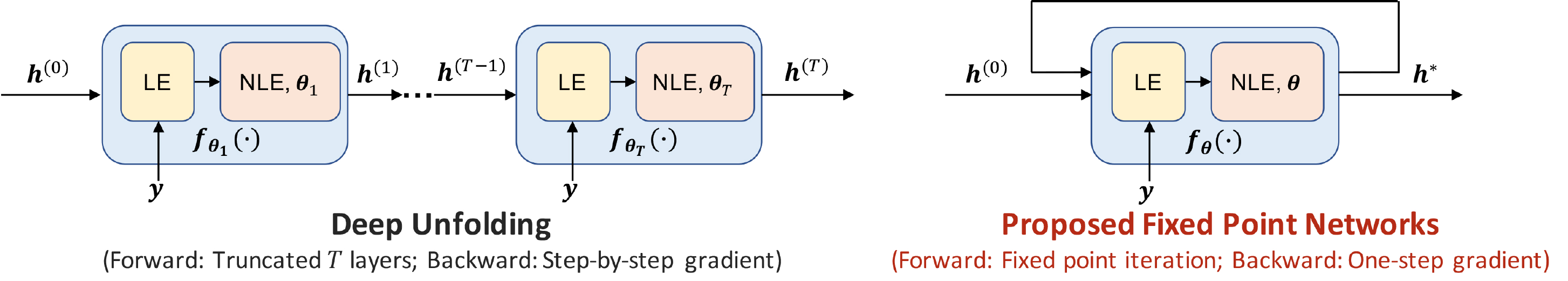}\caption{Comparison between the general frameworks of existing deep unfolding
methods and the proposed fixed point networks (FPNs). \label{fig:Comparison-between-the}}
\end{figure*}
\begin{equation}
\mathbf{\mathbf{y}=Mh+n}.\label{eq:real-inverse-1}
\end{equation}
The channel estimation aims to recover the channel representation
$\mathbf{h}$ from the measurement $\mathbf{y}$, with the knowledge
of the measurement matrix $\mathbf{M}$. If available, the statistics
of the noise $\mathbf{n}$ can also be utilized to enhance the performance.
In the following, we propose FPN-based algorithms to solve the problem.
Notice that (\ref{eq:real-inverse-1}) is also relevant to the general
linear inverse problems that are prevalent in the physical layer,
e.g., detection and decoding \cite{2019He}. The proposed method can
also provide inspirations for DL-based solutions to these problems. 

\section{FPNs for THz UM-MIMO Channel Estimation \label{sec:Fixed-Point-Networks:}}

\subsection{Preliminaries of Model-Driven DL\label{subsec:Preliminaries-of-Model-Driven}}

The measurement matrix $\mathbf{M}$ will be singular, if the number
of pilots is reduced. Prior information must be exploited in either
the form of regularization or prior distribution to ensure robust
estimation. The former formulates channel estimation as a regularized
LS (RLS) problem, i.e., $\min_{\mathbf{h}}\frac{1}{2}\|\mathbf{y}-\mathbf{M}\mathbf{h}\|_{2}^{2}+\lambda R(\mathbf{h})$,
where $\lambda$ is a positive scalar and $R(\mathbf{h})$ is a sparsity-inducing
regularizer, e.g., the $\ell_{1}$-norm. The RLS can be solved by
proximal gradient descent (PGD) or alternating direction method of
multipliers (ADMM) algorithms \cite{2011Boyd}. The latter formulates
channel estimation as maximum a posteriori (MAP) or minimum mean squared
error (MMSE) inference, which can be tackled by approximate message
passing (AMP) and OAMP \cite{2009Donoho,2017Ma}. Although these two
categories of algorithms have distinct principles and purposes, their
per-iteration update rules are similar \textit{in form}. In particular,
they can be divided into the linear estimator (LE) and the non-linear
estimator (NLE), as listed in Table \ref{tab:Per-iteration-update-rules}.
In the table, $\rho$ is the step size, the superscript $(t)$ is
the $t$-th iteration, $\text{Prox}(\cdot)$ denotes the proximal
operator \cite{2011Boyd}, $\mathbf{W}^{(t)}$ is the LE matrix of
OAMP, and $\eta_{t}(\cdot)$ denotes the NLE of OAMP. The last two
will be detailed later in Section III-D. The LE enforces the estimation
to be consistent with the received pilots, while the NLE ensures that
the estimation agrees with the prior knowledge of the channel. The
channel estimation algorithms can be interpreted as the \textit{fixed
point iteration} of the composition of the LE and the NLE, which can
run an \textit{arbitrary} number of iterations until convergence \cite{2019Bauschke}.
The estimated channel is then the stable state of the fixed point
iteration, called the \textit{fixed point }or equilibrium point. 

The LE relies on the received pilot signals, which is known perfectly,
and is easy to construct. However, the \textit{bottlenecks} of the
above algorithms all lie in the NLEs since their design requires prior
knowledge of the channel, which is difficult to acquire. Hand-crafted
regularizers or priors can only capture the rough information since
they are limited by their analytical structure. For example, the dictionary
matrix may not sparsify the channel well and can cause energy leakage
effects. Similarly, the empirically chosen base distribution to model
the prior can be mismatched with the true prior distribution. 

One natural idea is to replace NLEs with neural networks and then
learn to exploit the prior information of the channel from data, which
becomes well-motivated in theory due to two seminal works \cite{2013Venkatakrishnan,2016Metzler}.
In \cite{2013Venkatakrishnan}, the authors found that the proximal
operators in PGD and ADMM could be understood as MAP estimators of
Gaussian denoising problems. In \cite{2016Metzler}, the authors found
that the NLEs in AMP, OAMP and related algorithms can also be interpreted
as Gaussian noise denoisers. Deep neural networks, known as powerful
denoisers, are thus well-suited to substitute the NLEs, which leads
to the thriving of model-driven DL-based channel estimators \cite{2018He,2022He,2021Jin}. 

\subsection{Motivations of FPNs}

Model-driven DL-based, i.e., deep unfolding-based, estimators suffer
from several systematic drawbacks, which make them unsuitable for
THz UM-MIMO. In the sequel, we discuss these drawbacks and reveal
the motivations of our proposal. 

Deep unfolding-based estimators are constructed by first \textit{truncating}
a classical iterative algorithm to \textit{pre-defined} and \textit{fixed}
$T$ layers, and then substituting each NLE in layer $t$ by a deep
neural network parametrized by $\bm{\theta}_{t}$ \cite{2018He,2022He,2021Jin},
as illustrated in Fig. \ref{fig:Comparison-between-the}. In layer
$t$, the LE is denoted by $f_{\text{LE}}(\cdot;\mathbf{y})$ and
the NLE is given by $f_{\text{NLE},\bm{\theta}_{t}}(\cdot)$. The
overall update in layer $t$ is the composition of the LE and the
NLE, given by 
\begin{equation}
\mathbf{h}^{(t+1)}=f_{\bm{\theta}_{t}}(\mathbf{h}^{(t)};\mathbf{y})\triangleq(f_{\text{NLE},\bm{\theta}_{t}}\circ f_{\text{LE}})(\mathbf{h}^{(t)};\mathbf{y}).
\end{equation}
The difference between specific works only lies in the choice of the
base algorithm, which affects $f_{\text{LE}}(\cdot;\mathbf{y})$,
and the choice of the neural network, which impacts $f_{\text{NLE},\bm{\theta}_{t}}(\cdot)$.
The training process can be described by an optimization problem,
i.e.,
\begin{equation}
\begin{aligned}\min_{\bm{\Theta}=\{\bm{\theta}_{1},\ldots,\bm{\theta}_{T}\}} & \,\,\mathcal{L}(\mathbf{h}^{(T)};\mathbf{h}_{\text{gt}},\mathbf{y}),\\
\text{s.t.\ensuremath{\quad\quad}} & \,\,\mathbf{h}^{(T)}=(f_{\bm{\theta}_{T}}\circ\cdots\circ f_{\bm{\theta}_{1}})(\mathbf{h}^{(0)};\mathbf{y}),
\end{aligned}
\label{eq:Deep-unfolding}
\end{equation}
where $\mathcal{L}(\cdot;\cdot,\cdot)$ is the loss function, $\bm{\Theta}$
denotes the collection of trainable parameters, and $\mathbf{h}_{\text{gt}}$
denotes the ground-truth channel. The mapping $f_{\bm{\theta}_{T}}\circ\cdots\circ f_{\bm{\theta}_{1}}$
defines an \textit{explicit} neural network structure with \textit{truncated}
$T$ layers in forward propagation. 

Nevertheless, the accustomed formulation of deep unfolding can give
rise to several critical problems when it is applied to THz UM-MIMO
channel estimation. First, it scales poorly with the UM-MIMO array.
The required gradient in the training, i.e., backward, process of
(\ref{eq:Deep-unfolding}) is computed using the chain rule, i.e.,
\textit{step-by-step gradient}, which involves tracking and storing
all the intermediate states $\mathbf{h}^{(t)}$ and causes a high
space complexity of $\mathcal{O}(T)$ \cite{2019Bai,2022Fung}. For
UM-MIMO systems with thousands of antennas and complicated channel
conditions, the training cost is \textit{unaffordable}. Second, the
reliability is not guaranteed. Truncating the algorithm to $T$ layers
breaks the convergent nature of classical iterative algorithms. The
objective in (\ref{eq:Deep-unfolding}) only minimizes the estimation
accuracy of the final layer $\mathbf{h}^{(T)}$. Nonetheless, the
intermediate state $\mathbf{h}^{(t)}$ is not meaningful and tends
to oscillate frequently \cite{2022Chen-tianlong}. Third, the complexity
is high and not adaptive. Owing to the unreliable intermediate states,
deep unfolding algorithms are not adaptive and must be run for the
full $T$ layers, which causes excessive complexity. Lastly, the generalization
ability is poor in simulations, which cannot handle the changeable
channel conditions in THz UM-MIMO. 

Given these systematic drawbacks, it is important to rethink the feasibility
of the deep unfolding framework. Some previous works noticed the complexity
issue mentioned above, and proposed reinforcement learning-based modules
to realize early exit of the unfolding process \cite{2022Hu,2021Chen-JSAC}.
Nevertheless, the other issues still remain open, which motivates
us to propose a new and general framework to solve all these problems
and embed DL into iterative estimators in a theoretically sound manner. 

\subsection{General Ideas of FPNs}

As mentioned in Section \ref{subsec:Preliminaries-of-Model-Driven},
the estimated channel with classical iterative algorithms is the stable
state of the iteration, i.e., the \textit{fixed point} $\mathbf{h}^{*}$
defined by $\mathbf{h}^{*}=f_{\bm{\theta}}(\mathbf{h}^{*};\mathbf{y})$,
where the subscript $\bm{\theta}$ here indicates that the parameters
of the NLEs are the same in each iteration. If we can construct a
DL-involved mapping $f_{\bm{\theta}}(\cdot;\mathbf{y})$ whose repeated
application leads to a fixed point that corresponds to the estimated
channel, then the merits of classical algorithms will be perfectly
preserved. In addition, the powerful learning capability of deep neural
networks can be exploited at the same time, having the best of both
worlds. We refer to such framework as \textit{FPNs}, which is formulated
by
\begin{equation}
\begin{aligned}\underset{\bm{\theta}}{\min}\,\,\mathcal{L}(\mathbf{h}^{*};\mathbf{h}_{\text{gt}},\mathbf{y}),\,\,\,\,\,\,\,\,\,\, & \text{s.t.}\,\,\,\,\,\mathbf{h}^{*}=f_{\bm{\theta}}(\mathbf{h}^{*};\mathbf{y}).\end{aligned}
\label{eq:FPN-optimization}
\end{equation}
Let us first suppose that the fixed point $\mathbf{h}^{*}$ exists
and is also unique. We will discuss how to satisfy this assumption
soon. Then, (\ref{eq:FPN-optimization}) defines a \textit{bi-level}
optimization problem, where the inner level requires computing the
fixed point of the mapping $f_{\bm{\theta}}(\cdot;\mathbf{y})$, while
the outer level is the minimization of the loss with respect to the
parameters of the NLE neural network $\bm{\theta}$. This is drastically
different from the constraint in (\ref{eq:Deep-unfolding}) which
prescribes an \textit{explicit} neural network structure. On the contrary,
the constraint in (\ref{eq:FPN-optimization}) is rather defining
what one wants the DL-based mapping $f_{\bm{\theta}}(\cdot;\mathbf{y})$
to \textit{achieve}, other than providing an explicit structure, as
shown in Fig. \ref{fig:Comparison-between-the}. For example, to get
the fixed point $\mathbf{h}^{*}$, i.e., the estimated channel, given
different pilot signals $\mathbf{y}$, $f_{\bm{\theta}}(\cdot;\mathbf{y})$
may be executed for different numbers of times, with different computational
graphs. FPNs thus belong to \textit{implicit} DL \cite{2019Bai},
which can extend to an \textit{arbitrary} number of iterations until
convergence. 

To tackle the bi-level problem (\ref{eq:FPN-optimization}), one common
method is to compute the implicit gradient based on the implicit function
theorem \cite{2019Bai,2002Krantz}, as given in the following proposition. 
\begin{prop}
\label{prop:implicit-gradient}Given the fixed point equation $\mathbf{h}^{*}=f_{\bm{\theta}}(\mathbf{h}^{*};\mathbf{y})$
and the loss function $\mathcal{L}(\mathbf{h}^{*};\mathbf{h}_{\text{{\rm gt}}},\mathbf{y})$,
the gradient of the loss with respect to $\bm{\theta}$ is calculated
by 
\begin{equation}
\frac{\partial\mathcal{L}}{\partial\bm{\theta}}=\frac{\partial\mathcal{L}}{\partial\mathbf{h^{*}}}(\mathbf{I}-\frac{\partial f_{\bm{\theta}}(\mathbf{h}^{*};\mathbf{y})}{\partial\mathbf{h}^{*}})^{-1}\frac{\partial f_{\bm{\theta}}(\mathbf{h}^{*};\mathbf{y})}{\partial\bm{\theta}}.\label{eq:implicit-gradient}
\end{equation}
\end{prop}
\begin{IEEEproof}
Please refer to Appendix \ref{sec:Proof-implicit-gradient}. 
\end{IEEEproof}
From (\ref{eq:implicit-gradient}), we observe that calculating the
implicit gradient only requires the fixed point $\mathbf{h}^{*}$,
without the need of storing any intermediate states $\mathbf{h}^{(t)}$.
As a result, the memory complexity is only $\mathcal{O}(1)$ regardless
of the number of executed iterations, which is drastically smaller
in comparison to deep unfolding, i.e., $\mathcal{O}(T)$. However,
calculating the matrix inverse in (\ref{eq:implicit-gradient}) is
very costly given the high dimension of the channel. We turn to the
approximate gradient proposed in \cite{2022Fung} to reduce the computational
overhead, given by
\begin{equation}
\widehat{\left(\frac{\partial\mathcal{L}}{\partial\bm{\theta}}\right)}=\frac{\partial\mathcal{L}}{\partial\mathbf{h^{*}}}\frac{\partial f_{\bm{\theta}}(\mathbf{h}^{*};\mathbf{y})}{\partial\bm{\theta}}\thickapprox\frac{\partial\mathcal{L}}{\partial\bm{\theta}}.\label{eq:approximate-gradient}
\end{equation}
This has been proved to be a descending direction of the loss under
mild assumptions and achieved empirical success \cite{2022Fung}.
In addition, to realize (\ref{eq:approximate-gradient}) in the DL
libraries, e.g., Pytorch, one only needs to modify a few lines of
codes compared to the standard training procedure \cite[Section III]{2022Fung}.
We refer to this as \textit{one-step} gradient, as the backward process
only depends on one addition application of $f_{\bm{\theta}}(\cdot;\mathbf{y})$
at $\mathbf{h}^{*}$, regardless of how many iterations it takes to
reach the fixed point $\mathbf{h}^{*}$. 

With the low-cost training procedure at hand, the remaining problem
is how to ensure that the fixed point $\mathbf{h}^{*}=f_{\bm{\theta}}(\mathbf{h}^{*};\mathbf{y})$
exists and is unique, and how to find the fixed point efficiently.
Before going on, we first introduce two key concepts. 
\begin{defn}[Lipschitz continuity]
A mapping $f_{\bm{\theta}}(\cdot;\mathbf{y})$ is said Lipschitz
continuous if there exists a constant $L$ such that 
\[
\|f_{\bm{\theta}}(\mathbf{h}_{1};\mathbf{y})-f_{\bm{\theta}}(\mathbf{h}_{2};\mathbf{y})\|\leq L\|\mathbf{h}_{1}-\mathbf{h}_{2}\|
\]
holds for any $\mathbf{h}_{1},\mathbf{h}_{2}\in\text{dom}(f_{\bm{\theta}}(\cdot;\mathbf{y}))$. 
\end{defn}
\begin{defn}[Contraction]
A mapping $f_{\bm{\theta}}(\cdot;\mathbf{y})$ is a contraction mapping
if it is Lipschitz with constant $0\leq L<1$. 
\end{defn}
The existence of the fixed point and an efficient way to find it can
be ensured by fixed point theory \cite{2019Bauschke}. As long as
$f_{\bm{\theta}}(\cdot;\mathbf{y})$ is a contraction mapping (no
matter what detailed operations it contains), a simple repeated application
of $f_{\bm{\theta}}(\cdot;\mathbf{y})$ will make $\mathbf{h}^{(t)}$
converge in linear rate to the unique fixed point $\mathbf{h}^{*}$.
This can be formally stated in the following lemma. 
\begin{lem}[{Banach \cite[Theorem 1.50]{2019Bauschke}\label{Banach-Picard}}]
For any initial value $\mathbf{h}^{(0)}$, if the sequence $\{\mathbf{h}^{(t)}\}$
is generated via the relationship $\mathbf{h}^{(t+1)}=f_{\bm{\theta}}(\mathbf{h}^{(t)};\mathbf{y})$
and $f_{\bm{\theta}}(\cdot;\mathbf{y})$ is a contraction mapping
with Lipschitz constant $L$, then $\{\mathbf{h}^{(t)}\}$ converges
to the unique fixed point $\mathbf{h}^{*}$ of $f_{\bm{\theta}}(\cdot;\mathbf{y})$
with a linear convergence rate $L$. \label{lem:Banach}
\end{lem}
The above lemma indicates that if one can train a contraction $f_{\bm{\theta}}(\cdot;\mathbf{y})=(f_{\text{NLE},\bm{\theta}}\circ f_{\text{LE}})(\cdot;\mathbf{y})$
with DL-based components $\bm{\theta}$, then the convergence of FPNs
to the unique fixed point $\mathbf{h}^{*}$ in linear rate can be
theoretically guaranteed. That is to say, we should control the Lipschitz
constant of $f_{\bm{\theta}}(\cdot;\mathbf{y})$. Since the LEs of
classical iterative algorithms are all linear functions of $\mathbf{h}^{(t)}$,
their Lipschitz constants can be easily computed. Therefore, we only
need to control the Lipschitz constant of the neural network component
$f_{\text{NLE},\bm{\theta}}(\cdot)$. Given the following lemma, we
can work out the exact requirement of the neural network $\bm{\theta}$. 
\begin{lem}[\cite{2021Gouk}]
The composition of an $L_{1}$-Lipschitz and an $L_{2}$-Lipschitz
mapping is $L_{1}L_{2}$-Lipschitz. \label{lem:The-composition-of}
\end{lem}
The lemma above helps us identify the required Lipschitz constant
of neural network $f_{\text{NLE},\bm{\theta}}(\cdot)$ to ensure the
linear convergence. With such knowledge, we can control the Lipschitz
constant of $f_{\text{NLE},\bm{\theta}}(\cdot)$ during the training
process with many off-the-shelf methods \cite{2021Gouk}. Note that
Lipschitz-continuous neural networks can also contribute to the adversarial
robustness \cite{2021Gouk}, which is beneficial to the superb out-of-distribution
robustness of FPNs observed in the simulations in Section \ref{sec:Robust-Out-of-Distribution-Perfo}. 

In the sequel, we present an example of the FPN-enhanced iterative
channel estimator based on OAMP so as to illustrate the design guideline.
Similar procedures can also be applied to enhance other iterative
estimators, e.g., PGD and ADMM, which indicates the generality of
the proposed FPNs. 

\subsection{FPN-OAMP: Enhancing OAMP with FPNs}

OAMP is an efficient compressed sensing algorithm to solve channel
estimation problems, which consists of a de-correlated LE and a divergence-free
NLE, as shown in Table \ref{tab:Per-iteration-update-rules} \cite{2017Ma}.
In the sequel, we present the specific design of the FPN-enhanced
OAMP algorithm, i.e., FPN-OAMP. 

\subsubsection{Linear Estimator}

The LE of FPN-OAMP is similar to the original one in OAMP, given by
\begin{equation}
\mathbf{u}^{(t+1)}=f_{\text{LE}}(\mathbf{h}^{(t)};\mathbf{y})=\mathbf{h}^{(t)}+\mathbf{W}^{(t)}(\mathbf{y}-\mathbf{M}\mathbf{h}^{(t)})\label{eq:linear-estimator}
\end{equation}
where $\mathbf{W}^{(t)}$ is a de-correlated LE matrix. The matrix
$\mathbf{W}^{(t)}$ can be built upon the transpose and the pseudo-inverse
of $\mathbf{M}$, or the linear minimum mean square error (LMMSE)
matrix \cite{2017Ma}. The first two do not depend on the noise statistics
and are the same in each iteration, which match the FPN framework.
While the last one is the optimal form, it depends on the noise statistics
and requires computing a matrix inverse in each iteration, making
it too complicated for UM-MIMO systems. We choose the pseudo-inverse
LE due to its competitive performance and reasonable cost. The LE
matrix $\mathbf{W}^{(t)}$ is given by 
\begin{figure}[t]
\centering{}\includegraphics[width=6.5cm]{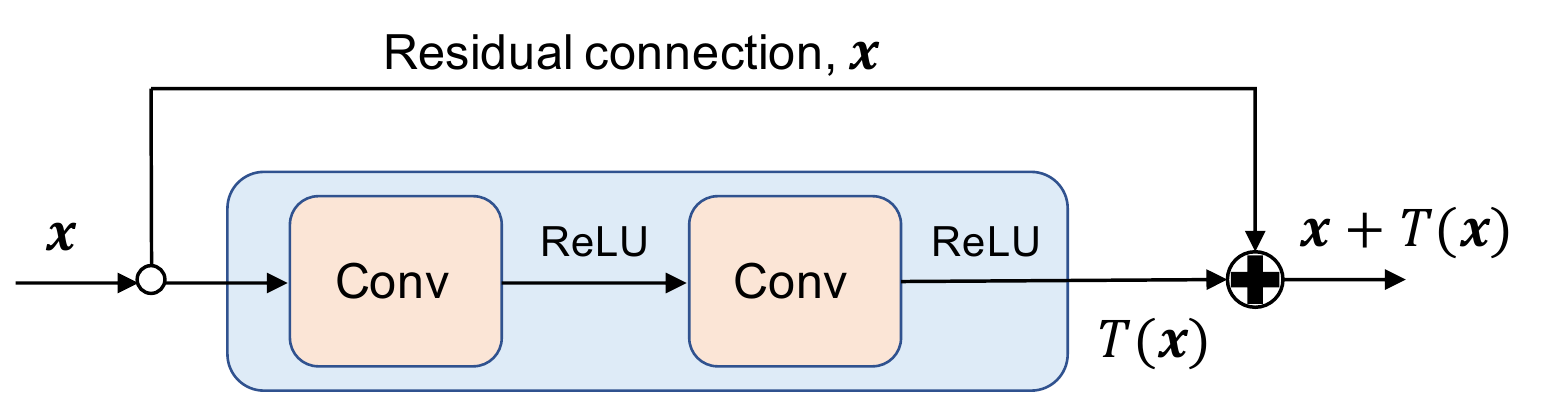}\caption{The structure of one RB. Let the input be $\mathbf{x}$, the output
is $\mathbf{x+}T(\mathbf{x})$. \label{fig:resnet-block}}
\end{figure}
\begin{equation}
\mathbf{W}^{(t)}=\eta\text{\ensuremath{\mathbf{M}^{\dagger}}}=\frac{2S\bar{S}}{\text{tr}(\mathbf{M}^{\dagger}\mathbf{M})}\mathbf{M}^{\dagger},
\end{equation}
where $\eta$ is the step size to guarantee that $\text{tr}(\mathbf{I}-\mathbf{W}\mathbf{M})=0$
holds, such that the LE is de-correlated. That is, the elements of
the NLE input error vector $\mathbf{u}^{(t+1)}-\mathbf{h}_{\text{gt}}$
are mutually uncorrelated with zero-mean and identical variances \cite{2017Ma}. 

\subsubsection{Non-linear Estimator\label{subsec:Non-linear-Estimator}}

The NLE in FPN-OAMP ensures that the estimation is consistent with
the prior knowledge of the channel. We replace the NLE in the original
OAMP with neural networks to learn to exploit the prior knowledge
from data. The NLE in FPN-OANP is given by 
\begin{figure*}[tbh]
\begin{equation}
\mathcal{L}_{\text{main}}(\mathbf{h}^{*};\mathbf{h}_{\text{gt}})=\frac{1}{n}\sum_{i=1}^{n}\left(\frac{\left\Vert \mathbf{h}_{\text{gt},i}-f_{\bm{\theta}}(\mathbf{h}_{i}^{*};\mathbf{y}_{i})\right\Vert _{1}}{\left\Vert \mathbf{h}_{\text{gt},i}\right\Vert _{1}}\right),\quad\mathcal{L}_{\text{aux}}(\mathbf{h}^{*};\mathbf{y})=\frac{1}{n}\sum_{i=1}^{n}\left(\frac{\left\Vert \mathbf{y}_{i}-\mathbf{M}f_{\bm{\theta}}(\mathbf{h}_{i}^{*};\mathbf{y}_{i})\right\Vert _{1}}{\left\Vert \mathbf{y}_{i}\right\Vert _{1}}\right).\label{eq:loss}
\end{equation}

\hrule
\end{figure*}
\begin{equation}
\mathbf{h}^{(t+1)}=\eta_{t}(\mathbf{u}^{(t+1)})=f_{\text{NLE},\bm{\theta}}(\mathbf{u}^{(t+1)}),
\end{equation}
which is a neural network parametrized by $\bm{\theta}$. Specifically,
$f_{\text{NLE},\bm{\theta}}(\cdot)$ is designed as a ResNet structure
with three residual blocks (RBs) \cite{2016He}. Before the RBs, $\mathbf{u}^{(t+1)}$
is first reshaped into a tensor form of $2S$ feature maps of size
$\sqrt{\bar{S}}\times\sqrt{\bar{S}}$, each corresponding to an SA,
and then passed through a convolution (Conv) layer to lift it to 64
feature maps. Each RB is formed by an identity mapping plus the compositions
of $3\times3$ Conv with 64 feature maps and ReLU activation function,
as shown in Fig. \ref{fig:resnet-block}. After the RBs, we apply
two $1\times1$ Conv layers with $2S$ feature maps, and reshape the
output into a vector with the same size as $\mathbf{u}^{(t+1)}$.
We adopt layer normalization in each RB for more stable training \cite{2016Ba},
which is suitable for recurrent neural networks, and so as the FPNs. 

There are two takeaways in the design of $f_{\text{NLE},\bm{\theta}}(\cdot)$.
First, since the planar AoSA is non-uniform, the relationship of the
channel between inter-SA and intra-SA antennas are different. Therefore,
one should separate the SA channels into different feature maps, other
than using the channel of the whole array as the input to the neural
network. Second, the residual links are important to make the NLE
approximately divergence-free. With such design, the statistical properties
of the original OAMP still hold well for FPN-OAMP \cite{2023Yu}.
By contrast, the NLE without residual links offers much worse performance. 

FPN-OAMP is summarized in \textbf{Algorithm 1}, which is designed
as the fixed point iteration of the contraction mapping $f_{\bm{\theta}}(\cdot;\mathbf{y})=(f_{\text{NLE},\bm{\theta}}\circ f_{\text{LE}})(\cdot;\mathbf{y})$.
We then discuss the key theoretical properties of FPN-OAMP and the
requirements on DL-based $f_{\text{NLE},\bm{\theta}}(\cdot)$ to ensure
that the mapping is a contraction. 

\begin{algorithm}[tbp]
\caption{FPN-OAMP for THz channel estimation}
\begin{algorithmic}[1]
\STATE {\bf Input:} Measurement matrix $\mathbf{M}$, received pilot signals $\mathbf{y}$, weights of the NLE $\bm{\theta}$, error tolerance $\epsilon$ \\
\STATE {\bf Output:} Estimated THz channel ${\mathbf{h}^*}$ (fixed point)\\
\STATE {\bf Initialize:} $\mathbf{h}^{(0)} \gets \mathbf{0}, t \gets 0$ \\
\STATE \textcolor{black}{Fixed point iteration of $f_{\bm{\theta}}(\cdot;\mathbf{y})$}:
\STATE {\bf while} $\|\mathbf{h}^{(t)}-f_{\bm{\theta}}(\mathbf{h}^{(t)};\mathbf{y})\|_2 > \epsilon$ {\bf do}
\STATE\hspace{\algorithmicindent} $\mathbf{h}^{(t+1)} \gets f_{\bm{\theta}}(\mathbf{h}^{(t)};\mathbf{y})$
\STATE\hspace{\algorithmicindent} $t \gets t+1$
\STATE ${\mathbf{h}^*} \gets \mathbf{h}^{(t)}$
\STATE {\bf return} ${\mathbf{h}^*}$  
\end{algorithmic}
\end{algorithm}

\subsection{Key Theoretical Properties }

\subsubsection{Representational Power}

The proposed FPNs restrict the parameters $\bm{\theta}$ of the DL-based
NLE $f_{\text{NLE},\bm{\theta}}(\cdot)$ to be identical in each iteration
in order to match the requirements of fixed point iteration. This
initially seems a drawback over existing deep unfolding methods, which
do not require the parameters to be identical and generally use a
different set of parameters for each layer, i.e., $\bm{\Theta}=\{\bm{\theta}_{1},\bm{\theta}_{2},\ldots,\bm{\theta}_{T}\}$.
Nevertheless, the representational power of FPNs has no notable loss
compared to deep unfolding due to the proposition below, even though
they are much cheaper to train \cite{2019Bai}. Simulations in Section
\ref{sec:Simulation-Results} also confirm the advantages of FPNs
over deep unfolding. 
\begin{prop}[{\cite[Theorem 3]{2019Bai}}]
For a $T$-layer deep unfolded network with different parameters
per layer, there exist FPNs that can represent the same network with
equivalent depth. 
\end{prop}

\subsubsection{Linear Convergence Rate}

We provide the proof for the linear convergence of FPN-OAMP to a unique
fixed point $\mathbf{h}^{*}$ and the requirement on the DL-based
NLE $f_{\text{NLE},\bm{\theta}}(\cdot)$. 
\begin{thm}
\label{thm:linear-convergence}The sequence $\{\mathbf{h}^{(t)}\}$
generated by FPN-OAMP updates $f_{\bm{\theta}}(\cdot;\mathbf{y})=(f_{\text{NLE},\bm{\theta}}\circ f_{\text{LE}})(\cdot;\mathbf{y})$
converges to a unique fixed point $\mathbf{h}^{*}$ with linear convergence
rate $L$ if the DL-based NLE $f_{\text{NLE},\bm{\theta}}(\cdot)$
is a contraction with Lipschitz constant $L$. 
\end{thm}
\begin{IEEEproof}
Please refer to Appendix \ref{sec:Proof-of-linear-convergence}. 
\end{IEEEproof}
We discuss how to train the DL-based NLE $f_{\text{NLE},\bm{\theta}}(\cdot)$
as a contraction mapping in Section \ref{subsec:Loss-Function-and}.
The linear convergence property ensures the reliability and efficiency
of FPN-OAMP, which is not available for existing deep unfolded methods.
In addition, since the inference process of FPNs is to find the fixed
point of a contraction, many advanced algorithms for this purpose,
e.g., Anderson acceleration \cite{2019Bauschke}, can be adopted to
potentially reach even super-linear convergence. 

\subsubsection{Adaptive Accuracy-Complexity Tradeoff}

As a corollary, the contraction property of FPN-OAMP further indicates
that the gap between $\mathbf{h}^{(t)}$ and $\mathbf{h}^{*}$ is
monotonically decreasing, according to the definition of Lipschitz
continuity. This indicates that intermediate states of FPN-OAMP will
be closer to the fixed point as the fixed point iteration goes. This
provides a user-defined tradeoff between complexity and accuracy,
which is valuable in practical deployment. 
\begin{cor}
Given the sequence $\{\mathbf{h}^{(t)}\}$ generated by FPN-OAMP updates
$f_{\bm{\theta}}(\cdot;\mathbf{y})$ with Lipschitz constant $L<1$
and fixed point $\mathbf{h}^{*}$, then $\left\Vert \mathbf{h}^{(t+1)}-\mathbf{h}^{*}\right\Vert _{2}\leq L\left\Vert \mathbf{h}^{(t)}-\mathbf{h}^{*}\right\Vert _{2}$
holds. \label{cor:adaptive-tradeoff}
\end{cor}

\subsection{Offline Training and Online Self-Adaptation\label{subsec:Loss-Function-and}}

The loss function we use, i.e., $\mathcal{L}(\mathbf{h}^{*};\mathbf{h}_{\text{gt}},\mathbf{y})$,
is the weighted sum of two different terms, given by
\begin{equation}
\mathcal{L}(\mathbf{h}^{*};\mathbf{h}_{\text{gt}},\mathbf{y})=\mathbf{\mathcal{L}_{\text{main}}(\mathbf{h}^{*};\mathbf{h}_{\text{gt}})}+\gamma\mathcal{L}_{\text{aux}}(\mathbf{h}^{*};\mathbf{y}),\label{eq:loss-function}
\end{equation}
where $\mathbf{\mathcal{L}_{\text{main}}(\mathbf{h}^{*};\mathbf{h}_{\text{gt}})}$
is the supervised main loss, $\mathcal{L}_{\text{aux}}(\mathbf{h}^{*};\mathbf{y})$
is the unsupervised auxiliary loss, and $\gamma$ is the hyper-parameter
balancing these two terms. For both the main and auxiliary loss functions
we utilize the normalized mean absolute error (NMAE) criteria, since
it results in better performance and robustness compared to the normalized
mean squared error (NMSE) loss according to the analysis in \cite{2020Qi}.
The expressions of the two terms are given in (\ref{eq:loss}) on
the top of the page. We let $\mathbf{h}^{*}$ and $\mathbf{h}_{\text{gt}}$
denote a batch of estimated/ground-truth channels, while $\mathbf{h}_{i}^{*}$
and $\mathbf{h}_{\text{gt},i}$ denote a specific sample in that batch. 

The offline training process is standard despite the use of the one-step
gradient instead of the step-by-step gradient of the chain rule. In
addition, we append a safeguarding step to ensure that the DL-based
NLE, i.e., $f_{\text{NLE},\bm{\theta}}(\cdot)$, in FPN-OAMP is contractive
by checking the approximate Lipschitz constant over the current batch
of data after each weight update, i.e.,
\begin{equation}
\hat{L}=\frac{{\scriptstyle {\displaystyle {\scriptstyle \sum_{i=1}^{n}\|f_{\text{NLE},\bm{\theta}}(\mathbf{h}_{i}^{*}\mathbf{+\bm{\delta}}_{i})-f_{\text{NLE},\bm{\theta}}(\mathbf{h}_{i}^{*})\|_{2}}}}}{{\scriptstyle \sum_{i=1}^{n}\|\mathbf{\bm{\delta}}_{i}\|_{2}}},
\end{equation}
where $\boldsymbol{\delta}_{i}$ denotes a small random perturbation.
If $\hat{L}>1$ is found, i.e., the contraction property does not
hold, we normalize the weight $\bm{\theta}$ according to $\hat{L}$,
similar to \cite{2022Yu}. Nevertheless, this is almost never violated
in our experiments, indicating that the training process itself encourages
the contraction property. 

Although FPN-OAMP can directly generalize to almost all the tested
distribution shifts in Section \ref{sec:Robust-Out-of-Distribution-Perfo},
it is still important to design an online self-adaptation scheme in
case that direct generalization fails. Our scheme is inspired by \cite{2022Darestani}
with two steps. First, include an unsupervised auxiliary loss $\mathcal{L}_{\text{aux}}(\mathbf{h}^{*};\mathbf{y})$
at the offline training stage. Second, at the online deployment stage,
if potential performance drop is detected\footnote{In our recent work \cite{2023Yu}, we propose a preliminary method
to realize this goal. However, detailed discussion on this is out
of the scope of this paper. }, fine-tune the model based on the offline-trained parameters using
the auxiliary loss for the one particular received pilot signal $\mathbf{y}$.
The overhead of one iteration of the online fine-tuning is roughly
equal to doing one additional forward propagation, which is quite
cheap. In practice, we find that around 5 iterations of fine-tuning
is often enough. Notice that the online self-adaptation is only a
backup option, since direct generalization can already handle most
cases of distribution shifts. 

\subsection{Complexity Analysis}

We analyze the complexity based on the real-valued system model (\ref{eq:real-inverse-1}).
The complexity of the LE in FPN-OAMP is dominated by matrix-vector
product, given by $\mathcal{O}(4S^{2}\bar{S}Q)$, because the LE matrix
$\mathbf{W}^{(t)}$ is the same in each iteration and can be pre-computed
and cached\footnote{Note that this does not mean the algorithms can only work with the
pre-specified measurement matrix. In Section \ref{sec:Robust-Out-of-Distribution-Perfo},
we will show that the proposed FPN-OAMP trained with a single measurement
matrix can directly generalize to a wide variety of measurement matrix
conditions.}. The complexity of the NLE in FPN-OAMP depends on the number of floating
point operations (FLOPs) of the neural network, which is constant
and denoted by $c$. To reach an $\epsilon$-optimal precision of
the fixed point, FPN-OAMP requires only $\mathcal{O}(\log\frac{1}{\epsilon})$
iterations due to linear convergence. The overall complexity is $\mathcal{O}(\log\frac{1}{\epsilon}(4S^{2}\bar{S}Q+c))$,
which scales linearly with the number of antennas. To illustrate the
complexity straightforwardly, we provide the running time in Section
\ref{sec:Simulation-Results}. 

\section{\textcolor{black}{Extension to Wideband Systems}}

In previous sections, we adopted the narrowband system model to better
illustrate the ideas of the proposed algorithms. Nevertheless, in
practice, to exploit the huge available bandwidth, THz UM-MIMO systems
are likely to be operated in the wideband mode. We discuss how our
proposed method can be readily extended in the following. 

\textcolor{black}{Consider a wideband THz UM-MIMO orthogonal frequency
division multiplexing (OFDM) system}\footnote{\textcolor{black}{Another promising alternative to OFDM is single-carrier
frequency domain equalization \cite{2021Dovelos}, to which the proposed
method can also be easily extended. }}\textcolor{black}{{} with the same BS array as in Section \ref{sec:System-Model-and}.
We consider $K$ subcarriers over a bandwidth of $B$ at the center
frequency $f_{c}$. For an arbitrary UE, the real-valued equivalent
of the received signal at the $k$-th subcarrier, $\mathbf{y}[k]\in\mathbb{R}^{2SQ\times1}$,
can be obtained in a similar manner as the narrowband case based on
(\ref{eq:real-inverse-1}), i.e., 
\begin{equation}
\mathbf{y}[k]=\mathbf{M}\mathbf{h}[k]+\mathbf{n}[k],\label{eq:wideband-real-inverse}
\end{equation}
where $k=1,2,\ldots,K$ is the subcarrier index at frequency $f_{k}=f_{c}+(k-1-\frac{K-1}{2})\frac{B}{K}$,
and $\mathbf{M}\in\mathbb{R}^{2SQ\times2S\bar{S}}$ is the measurement
matrix defined in Section \ref{sec:System-Model-and}. It is irrelevant
to the subcarrier index $k$ since the analog combiner is shared across
different subcarriers \cite{2020Zhang}. Similar to $\mathbf{h}$
in the narrowband case, the channel vector at the $k$-th subcarrier,
i.e., $\mathbf{h}[k]\in\mathbb{R}^{2S\bar{S}\times1}$, can be generated
by replacing $f_{c}$ with $f_{k}$ in (\ref{eq:channel}). However,
the gaps among the carrier frequencies $f_{k}$ are fairly large owing
to the huge available bandwidth $B$ at the THz band, making the array
response $\mathbf{a}(\phi_{l},\theta_{l},r_{l},f_{k})$ fairly frequency-selective.
Even for the same multipath component, the beam power can still vary
considerably at different subcarriers, which leads to the }\textit{\textcolor{black}{spatial
wideband effect}}\textcolor{black}{, or }\textit{\textcolor{black}{beam
squint effect}}\textcolor{black}{{} \cite{2018WangTSP}. As a result,
the combining gain or the effective signal-to-noise-ratio (SNR) is
also unequal across subcarriers given that the analog combiner is
shared \cite{2021Dovelos}. This renders a key difference between
narrowband and wideband channels. }

\textcolor{black}{We discuss two possible approaches to extending
the algorithmic framework to wideband THz systems. }

\subsection{\textcolor{black}{Narrowband Dataset}}

\textcolor{black}{The most direct way is to employ the narrowband
FPN-OAMP algorithm (trained for the central frequency $f_{c}$) to
solve the channel estimation problem at each subcarrier $k$ in a
parallel manner. For practical implementation, this can be easily
achieved by increasing the testing batch size to the number of subcarriers.
This method directly reuses the narrowband estimator without retraining
thanks to the strong generalization capability. This method ignores
the correlation between different subcarriers. Nevertheless, the performance
is still competitive thanks to the strong generalization ability. }

\subsection{\textcolor{black}{Wideband Dataset}}

\textcolor{black}{The second method also deals with the wideband channel
estimation problem through $K$ parallel subproblems. The key difference
is that the network is trained using the wideband dataset, which covers
all subcarriers, by treating the narrowband channel at each subcarrier
as a separate training sample. After training, the inference procedure
is the same as above. This can exploit the correlation among different
subcarriers during training to tackle beam squint. }

\section{Simulation Results \label{sec:Simulation-Results}}

\subsection{\textcolor{black}{Simulation Setup \label{subsec:Simulation-Setups}}}

\textcolor{black}{We first consider the narrowband systems, while
extension to the wideband systems is discussed at the end of this
section. The }main system parameters and their values are listed in
Table \ref{tab:Key-Simulation-Parameters} \cite{2021Dovelos}. To
model the hybrid-field channel conditions, the scatterer distance
$r_{l}$ is set as a uniformly distributed random variable, spanning
both the far-field and near-field regions. We adopt the NMSE between
the estimated and the ground-truth channel as the performance metric,
which is averaged over the testing dataset. The following six benchmarks
are compared: 
\begin{table}[t]
\caption{Key system parameters\label{tab:Key-Simulation-Parameters}}

\centering{}%
\begin{tabular}{l|l}
\hline 
\textbf{Parameter} & \textbf{Value}\tabularnewline
\hline 
Number of SAs / RF chains & $S=4$\tabularnewline
Number of AEs per SA & $\bar{S}=256$\tabularnewline
Total number of BS antennas & $S\bar{S}=1024$\tabularnewline
Carrier frequency & $f_{c}=300$ GHz\tabularnewline
AE spacing & $d_{a}=\lambda_{c}/2=0.0005$ m\tabularnewline
SA spacing & $d_{\text{sub}}=56\lambda_{c}=0.056$ m\tabularnewline
Pilot length & $Q=128$\tabularnewline
Under-sampling ratio & $\rho=\frac{SQ}{S\bar{S}}=50\%$\tabularnewline
Azimuth AoA & $\theta_{l}\sim\mathcal{U}(-\pi/2,\pi/2)$\tabularnewline
Elevation AoA & $\phi_{l}\sim\mathcal{U}(-\pi,\pi)$\tabularnewline
Angle of incidence & $\varphi_{\text{in},l}\sim\mathcal{U}(0,\pi/2)$\tabularnewline
Number of paths & $L=5$\tabularnewline
Rayleigh distance & $D_{\text{Rayleigh}}=20$ m\tabularnewline
LoS path length & $\text{\ensuremath{r_{1}=30}}$ m\tabularnewline
NLoS scatterer distance ($l>1$) & $r_{l}\sim\mathcal{U}(10,25)$ m\tabularnewline
Time delay of LoS path & $\tau_{1}=100$ nsec\tabularnewline
Time delay of NLoS paths ($l>1$) & $\tau_{l}\sim\mathcal{U}(100,110)$ nsec\tabularnewline
Molecular absorption coefficient & $k_{\text{abs}}=0.0033$ m$^{-1}$\tabularnewline
Refractive index & $n_{t}=2.24-j0.025$\tabularnewline
Roughness factor & $\sigma_{\text{rough}}=8.8\times10^{-5}$ m\tabularnewline
\hline 
\end{tabular}
\end{table}

\begin{itemize}
\item \textbf{LS}: Least squares estimation. 
\item \textbf{OMP}: Sparse reconstruction-based method with the orthogonal
matching pursuit algorithm \cite{2021Dovelos,2022Cui}. 
\item \textbf{OAMP}: Bayesian estimation via OAMP algorithm with the LMMSE
LE, and Bernoulli-Gaussian prior \cite{2017Ma}. 
\item \textbf{FISTA}: Sparse reconstruction-based method with the fast iterative
soft thresholding algorithm \cite{2009Beck}. 
\item \textbf{EM-GEC}: Bayesian estimation via EM-assisted generalized expectation
consistent signal recovery with Gaussian mixture prior \cite{2019Wang}. 
\item \textbf{ISTA-Net+}: state-of-the-art deep unfolding method based on
the iterative soft thresholding algorithm \cite{2018Zhang,2021Jin}.
\begin{figure*}[tbh]
\begin{minipage}[t]{0.32\textwidth}%
\begin{center}
\includegraphics[width=6cm]{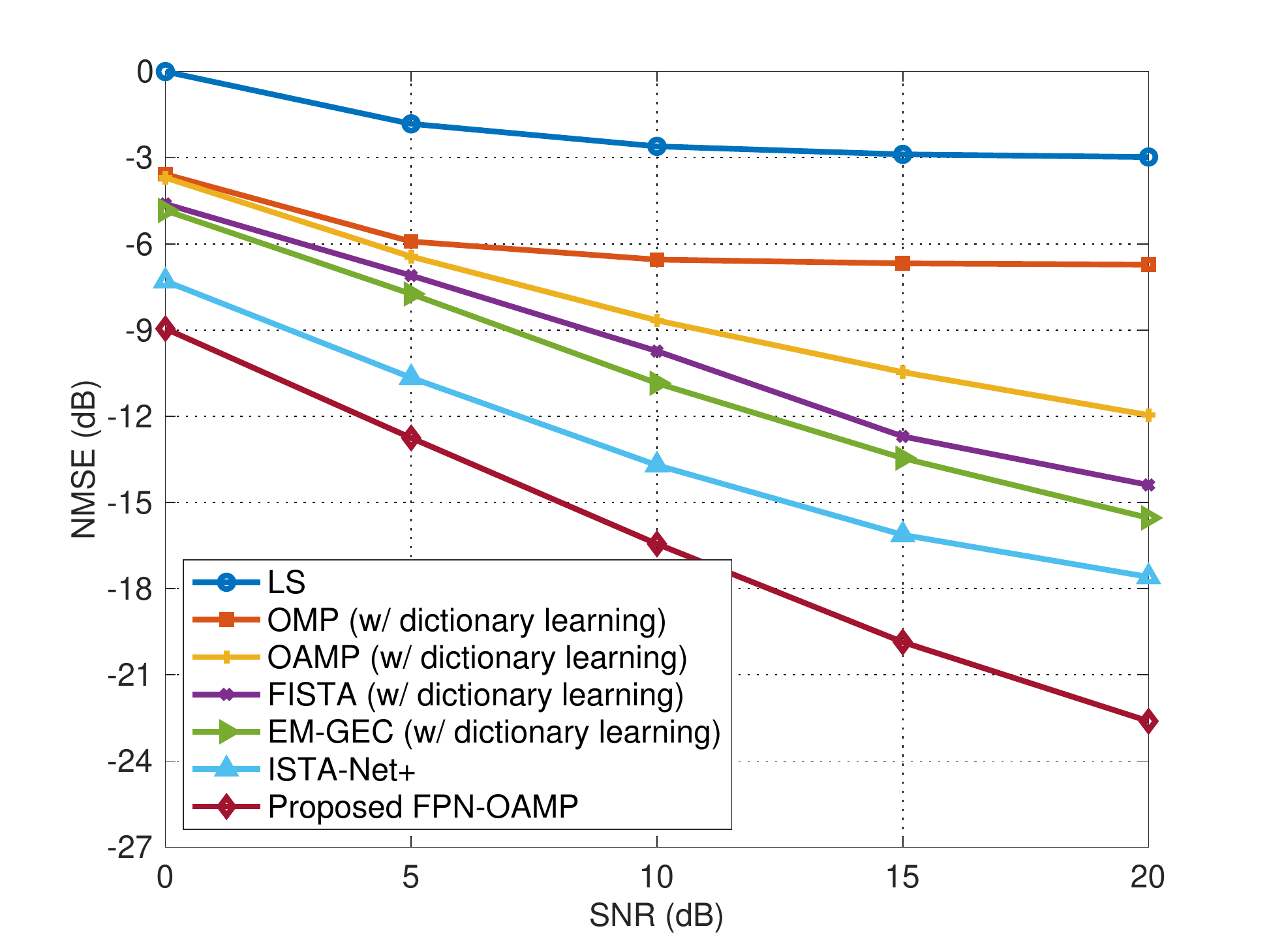}\caption{NMSE comparison at different SNR levels. \label{fig:NMSE-performance-comparison}}
\par\end{center}%
\end{minipage}\enskip{}%
\begin{minipage}[t]{0.32\textwidth}%
\begin{center}
\includegraphics[width=6cm]{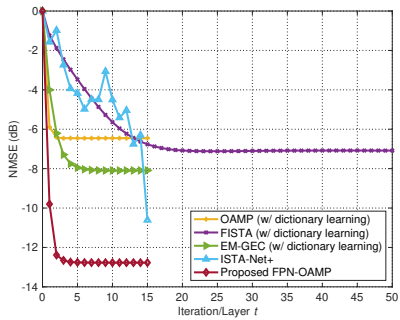}\caption{NMSE at iteration/layer $t$ ($\text{SNR}=5$ dB). \label{fig:NMSE-iteration-5dB}}
\par\end{center}%
\end{minipage}\enskip{}%
\begin{minipage}[t]{0.32\textwidth}%
\begin{center}
\includegraphics[width=6cm]{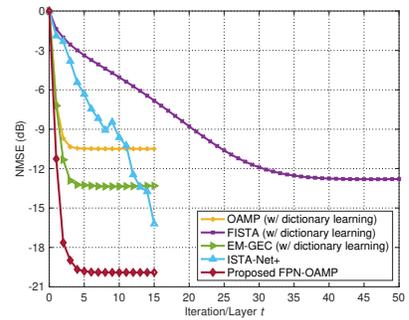}\caption{NMSE at iteration/layer $t$ ($\text{SNR}=15$ dB). \label{fig:NMSE-iteration-15dB}}
\par\end{center}%
\end{minipage}
\end{figure*}
\end{itemize}
Both ISTA-Net+ and the proposed FPN-OAMP are implemented by using
Pytorch and trained for 100 epochs using the Adam optimizer. We set
$\gamma=0.3$ in (\ref{eq:loss-function}), and use a batch size of
128 and an initial learning rate of 0.001. The learning rate is reduced
by half after every 30 epochs. The training, validation, and testing
datasets respectively consist of 80000, 5000, and 5000 samples. The
SNR levels of the training and validation samples are randomly drawn
from 0 to 20 dB. We observe that mixed-SNR training only causes a
small drop in performance compared to dedicated-SNR training. Therefore,
only a \textit{single} set of parameters is trained. Due to the limitation
of deep unfolding methods, ISTA-Net+ should be truncated to a \textit{finite}
number of layers. When training and testing ISTA-Net+, we\textit{
}set this number\textit{ }as 15, since a further increase can offer
only negligible gain. When training the FPN-OAMP, we set the error
tolerance $\epsilon$ as 0.01 and the maximum number of iterations
as 15 for fair comparison with ISTA-Net+. At the testing stage, one
can run FPN-OAMP for an \textit{arbitrary} number of iterations until
convergence. The training of FPN-OAMP takes only 50 minutes to complete
and consumes less than 1.5 GB of memory on Nvidia A40 GPU. 

We then introduce the choice of the dictionary $\mathbf{F}$. For
deep learning based methods (i.e., ISTA-Net+ and FPN-OAMP), we adopt
the DFT-based dictionary as introduced in Section III. For other benchmarks
where the quality of the sparse representation is important, we resort
to dictionary learning to enhance their performance \cite{2018Ding}.
The dictionary $\mathbf{F}$ is obtained by an $\ell1$-sparse coding
problem over the training dataset, i.e., 
\begin{equation}
\min_{\mathbf{F},\text{\ensuremath{\mathbf{h}_{1},\mathbf{h}_{2},\ldots,\mathbf{h}_{n}}}}\frac{1}{n}\sum_{i=1}^{n}(\frac{1}{2}\|\mathbf{\mathbf{\check{h}}}_{i}-\mathbf{Fh}_{i}\|_{2}^{2}+\lambda\|\mathbf{h}_{i}\|_{1}),\label{eq:dic-learn}
\end{equation}
where $\mathcal{C}$ is the constraint set of the dictionary, i.e.,
$\mathcal{C}\triangleq\{\mathbf{F}\in\mathbb{R}^{2S\bar{S}\times2S\bar{S}},\text{s.t. }\|(\mathbf{F})_{:,j}\|_{2}^{2}\leq1,\forall j=1,2,\ldots,2S\bar{S}\}$,
$n$ is the number of samples, and $\lambda$ is a hyper-parameter
\cite{2018Ding}. In the objective, $\mathbf{\mathbf{\check{h}}}_{i}=[\text{\ensuremath{\Re}}(\mathbf{\mathbf{\tilde{h}}}_{i})^{T},\text{\ensuremath{\Im}}(\mathbf{\mathbf{\tilde{h}}}_{i})^{T}]^{T}\in\mathbb{R}^{2S\bar{S}\times1}$
is the real-valued equivalent of the spatial channel $\mathbf{\mathbf{\tilde{h}}}_{i}\in\mathbb{C}^{S\bar{S}\times1}$.
To solve the problem efficiently, we use the algorithm proposed in
\cite{2009Mairal}. 

\subsection{Superior In-Distribution Performance}

In Fig. \ref{fig:NMSE-performance-comparison}, we present the NMSE
comparison at different SNR levels. It demonstrates that our proposed
FPN-OAMP outperforms all five benchmarks by a large margin. Compared
with its base algorithm OAMP, the performance gain of FPN-OAMP is
as large as 10 dB. This indicates that the CNN components of FPN-OAMP
can effectively extract and exploit the structures of the complicated
hybrid-field THz UM-MIMO channel. It is worth noting that, although
we have augmented the CS-based benchmarks with dictionary learning,
their performance still has a notable gap compared to DL-based ones.
In addition, FPN-OAMP always outperforms the deep unfolding method,
ISTA-Net+.
\begin{figure*}[tbh]
\begin{minipage}[t]{0.32\textwidth}%
\begin{center}
\includegraphics[width=6cm]{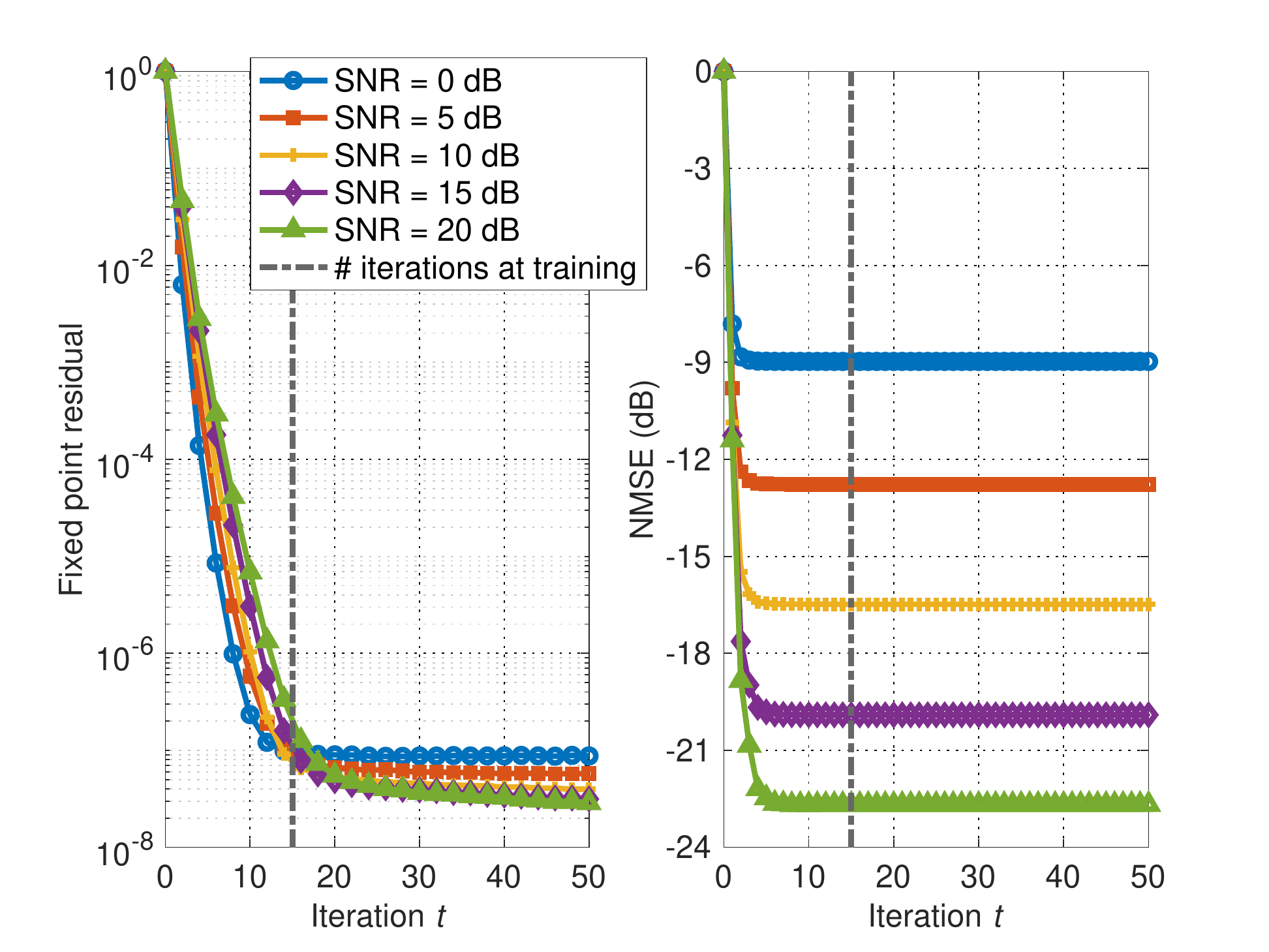}\caption{Convergence of the FPN-OAMP in terms of fixed point residual (left)
and NMSE (right). \label{fig:Convergence-of-the}}
\par\end{center}%
\end{minipage}\enskip{}%
\begin{minipage}[t]{0.32\textwidth}%
\begin{center}
\includegraphics[width=6cm]{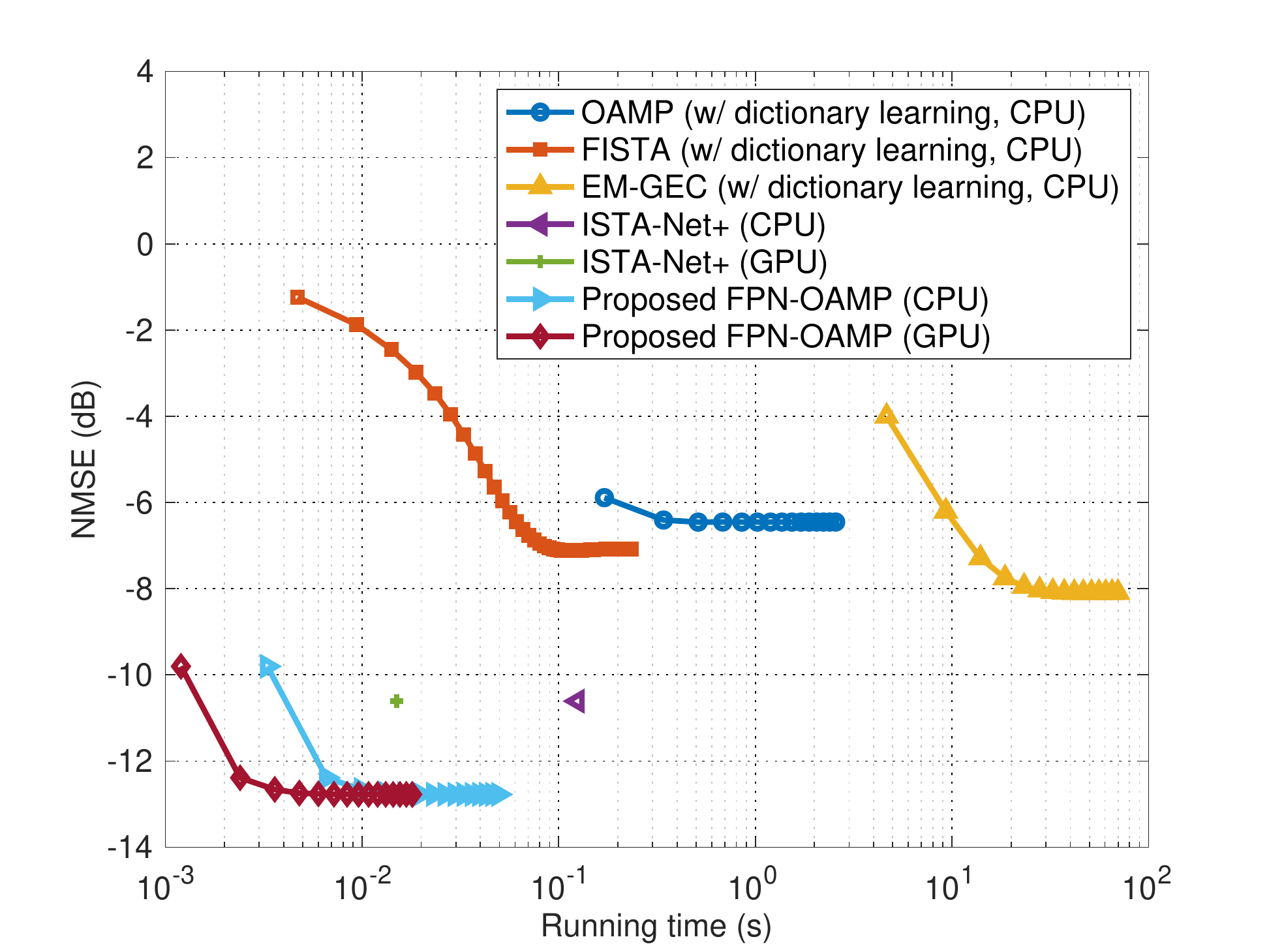}\caption{NMSE versus running time ($\text{SNR}=5$ dB).\label{fig:computational-complexity}}
\par\end{center}%
\end{minipage}\enskip{}%
\begin{minipage}[t]{0.32\textwidth}%
\begin{center}
\includegraphics[width=6cm]{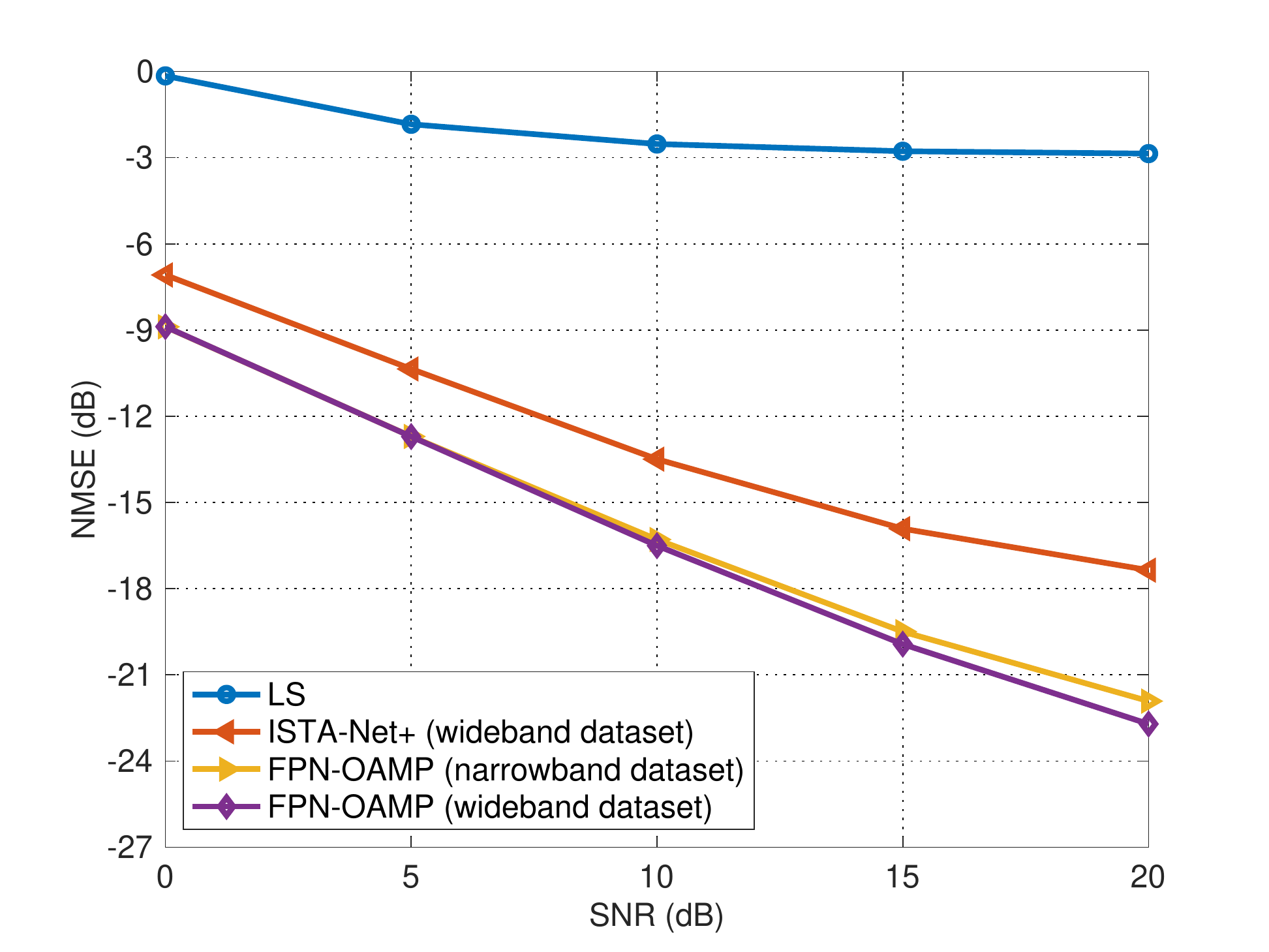}\caption{NMSE comparison at different SNR levels in wideband THz UM-MIMO systems.
\label{fig:NMSE-performance-comparison-wideband}}
\par\end{center}%
\end{minipage}
\end{figure*}

In Figs. \ref{fig:NMSE-iteration-5dB} and \ref{fig:NMSE-iteration-15dB},
we compare the NMSE evaluated at different numbers of iterations/layers\footnote{The term\textsl{ }\textit{iteration} is used when the algorithm we
refer to can extend to an arbitrary depth, e.g., FPN-OAMP. By contrast,
the term \textit{layer} is adopted when the algorithm is truncated
to a pre-defined and fixed depth, e.g., ISTA-Net+. }, when the SNR is 5 dB and 15 dB, respectively. LS and OMP are not
plotted since they do not produce intermediate results. In both cases,
the proposed FPN-OAMP converges rapidly within 5 iterations. Furthermore,
the performance of FPN-OAMP at the \textit{second} iteration is already
better than the final performance of all benchmarks. It is also observed
that the accuracy of all methods, except ISTA-Net+, increases consistently
with the number of iterations. The instability of ISTA-Net+ is mainly
because it is truncated to a fixed number of layers, and only the
final performance is optimized during training, with no control of
the intermediate states. Due to such instability, to maintain competitive
performance, ISTA-Net+ must use a fixed number of layers during both
training and testing. By contrast, the proposed FPN-OAMP is adaptive
in the number of iterations and has much more stable performance. 

\subsection{Linear Convergence Rate}

In Fig. \ref{fig:Convergence-of-the}, we evaluate the convergence
of the proposed FPN-OAMP in terms of fixed point residual (left) and
NMSE (right). Fixed point residual is given by $\mathbb{E}\{\|\mathbf{h}^{(t)}-\mathbf{h}^{*}\|/\|\mathbf{h}^{*}\|\}$.
In the left figure, the fixed point $\mathbf{h}^{*}$ is obtained
by running FPN-OAMP for 200 iterations, well beyond that during training,
i.e., 15 iterations. The curves are all linear before reaching convergence,
which verifies the \textit{linear convergence} rate proved in \textbf{Theorem
\ref{thm:linear-convergence}}, and also justifies the safeguarding
strategy in Section \ref{subsec:Loss-Function-and}. We also observe
that the fixed point residual is \textit{monotonically} decreasing,
which agrees with \textbf{Corollary \ref{cor:adaptive-tradeoff}}.
This eliminates the need of tricky stopping criteria. The difference
norm $\epsilon$ or the running time budget can serve as good stopping
criteria with theoretical support. In the right plot, it is observed
that the NMSE converges after 2-5 iterations. The number of iterations
required is slightly larger at higher SNR levels. We run the algorithm
for a much larger number of iterations compared with the training
stage, to confirm that it can extend to an arbitrary number of iterations. 

\subsection{Adaptive Accuracy-Complexity Tradeoff}

In Fig. \ref{fig:computational-complexity}, we compare the NMSE versus
the running time when the SNR is 5 dB. The CPU running time is tested
on Intel Core i7-9750H, while the GPU time is tested on Nvidia A40
GPU. The recorded running time contains only the online inference
stage, while the time consumption of the offline stage, e.g., dictionary
learning and DL training, is not included. As can be seen, the proposed
FPN-OAMP always costs the least time to converge \textemdash{} only
a few milliseconds even when tested on CPU. The per-iteration running
time of FPN-OAMP is as low as that of the first-order optimization
method FISTA, but it converges much faster with far better performance.
Additionally, it is observed that, for any given running time budget,
FPN-OAMP can always achieve a significantly better performance compared
with all benchmarks. Furthermore, unlike deep unfolding methods, e.g.,
ISTA-Net+, which requires a \textit{fixed} number of layers and therefore
a \textit{fixed} time budget, the running time of FPN-OAMP is \textit{adaptive}.
By adjusting the error tolerance $\epsilon$ or the running time budget,
a user-defined tradeoff between complexity and accuracy can be readily
achieved. This unique advantage is very important since the latency
requirement and the computational capability may often vary in practical
deployment. 

\subsection{\textcolor{black}{Extension to Wideband Systems}}

\textcolor{black}{In Fig. \ref{fig:NMSE-performance-comparison-wideband},
we compare the NMSE performance at different SNR levels for the wideband
case. Besides the settings in Section \ref{subsec:Simulation-Setups},
the wideband system utilizes OFDM modulation with bandwidth $B=15$
GHz and $K=32$ subcarriers \cite{2021Dovelos}. We adopt the HITRAN
database to generate the frequency-dependent molecular absorption
loss $k_{\text{abs}}$ \cite{2021Tarboush,2017Gordon}. For FPN-OAMP
with the narrowband dataset, we directly utilize the same narrowband
channel estimator at the center frequency $f_{c}$. For the wideband
dataset-based approaches, we train the network using the same mixed-SNR
training strategy as in Section \ref{subsec:Simulation-Setups}. Simulation
results show that both the narrowband and wideband variants of FPN-OAMP
can significantly outperform ISTA-Net+. In addition, we observe that
the narrowband variant is directly applicable to the wideband systems
without notable performance loss, which demonstrates the strong generalization
capability of FPN-OAMP over different frequency bands, even with the
existence of beam squint. }

\section{Robust Out-of-Distribution Performance\label{sec:Robust-Out-of-Distribution-Perfo}}

Distribution shifts, i.e., the case where the distributions of the
training and testing data differ, are prevalent in practical deployments
and may cause serious performance degradations for DL-based channel
estimators. This is the notorious \textit{out-of-distribution (OoD)}
problem. The distribution shifts related to channel estimation can
be grouped as 
\begin{itemize}
\item Noise distribution shifts in $\mathbf{n}$, 
\item Channel distribution shifts in $\mathbf{h}$, 
\item Measurement distribution shifts in $\mathbf{M}$. 
\end{itemize}
With extensive simulations, we show the strong generalization capability
of FPN-OAMP to all these distribution shifts. For the rare cases where
direct generalization fails, we verify the effectiveness of the online
self-adaptation scheme. 

In the sequel, the source distribution \textit{by default} refers
to the same simulation setup described in Section \ref{subsec:Simulation-Setups}.
In the tables below, we only list the particular configuration that
is different in the source and target distributions. The rest configurations
are kept unchanged. By \textit{in-distribution}, we refer to the case
where the model is trained and tested both on the target distribution.
The training procedure on the target distribution dataset is exactly
the same as that described in Section \ref{subsec:Simulation-Setups}.
By \textit{OoD}, we mean that the model is trained on the source distribution
but tested on the target distribution. The performance is also averaged
over 5000 testing samples. 

\subsection{Noise Distribution Shifts}

In the first two rows of Table \ref{tab:OoD-noise}, we study the
influence of noise level shifts. For noise levels that are either
lower or higher than the training configuration, the OoD performance
is close to the in-distribution one. This demonstrates that the proposed
FPN-OAMP is robust in face of noise level shifts, although it does
not use any information of the noise statistics. Besides, we can further
reduce the performance loss by enlarging the training SNR range based
on practical needs. 

In the third row of Table \ref{tab:OoD-noise}, we examine the noise
type shifts. We adopt the $\alpha$-stable distribution to model the
impulsive noise \cite{1995Nikias,2022Ma}, which is defined by the
stability parameter $0<\alpha\leq2$, the skewness parameter $-1\leq\beta\leq1$,
and the dispersion parameter $\gamma>0$. Since the $\alpha$-stable
noise has no limited variance, the normal SNR definition becomes invalid.
We alternatively use the generalized SNR (GSNR) defined as the ratio
of the signal power and the dispersion parameter $\gamma$ \cite{1995Nikias}.
For evaluation, we let $\alpha=1.7$, $\beta=0.2$, and set $\gamma$
as such that the GSNR equals 15 dB. The result shows that the FPN-OAMP
model trained with AWGN can directly generalize to the impulsive noise
case. 

\subsection{Channel Distribution Shifts}

In Table \ref{tab:OoD-channel}, we present the OoD generalization
performance of the proposed FPN-OAMP under channel distribution shifts.
The results in this table are all tested when the SNR is 15 dB, and
the noise distribution type is AWGN. 

In the first row, we consider the influence of LoS blockage, which
may frequently occur in THz UM-MIMO systems due to the high penetration
loss \cite{2021Sarieddeen}. The channels in the source distribution
all consist of one LoS and four NLoS paths, while the channels in
the target distribution only consist of four NLoS paths, where the
NLoS scatterer distance $r_{l}$ follows $r_{l}\sim\mathcal{U}(10,25)$
m. The result shows that the performance drop is less than 0.2 dB,
suggesting that LoS blockage almost has no negative effect on our
proposed FPN-OAMP. 

In the second and third rows, we check the effect of the number of
paths. The results suggest that breaking away from the original number
of paths will cause nearly no detriment to the performance of FPN-OAMP. 

In the fourth and fifth rows, we study the influence of field mismatch.
To model the near-field only channel, we set the source/scatterer
distance as $r_{l}\sim\mathcal{U}(10,20)$ m, which is within the
Rayleigh distance. We instead set the distance beyond the Rayleigh
distance, as $r_{l}\sim\mathcal{U}(20,30)$ m, to model the far-field
only channel. The performance drop is smaller than 0.1 dB, again demonstrating
the robustness of FPN-OAMP. 
\begin{table}[t]
\begin{centering}
\caption{OoD generalization under noise distribution shifts \label{tab:OoD-noise}}
\par\end{centering}
\centering{}%
\begin{tabular}{c|c|c|c}
\hline 
\textbf{Source} & \textbf{Target} & \textbf{In-distribution} & \textbf{OoD NMSE}\tabularnewline
\textbf{distribution} & \textbf{distribution} & \textbf{NMSE} & \textbf{(w/o self-ada.)}\tabularnewline
\hline 
$\text{SNR}\in[0,20]$ dB & $\text{SNR}=-5$ dB & $-$5.44 dB & $-$5.34 dB\tabularnewline
$\text{SNR}\in[0,20]$ dB & $\text{SNR}=25$ dB & $-$24.85 dB & $-$24.27 dB\tabularnewline
AWGN & Impulsive noise & $-$19.97 dB & $-$19.57 dB\tabularnewline
\hline 
\end{tabular}
\end{table}
\begin{table}[t]
\begin{centering}
\caption{OoD generalization under channel distribution shifts \label{tab:OoD-channel}}
\par\end{centering}
\centering{}%
\begin{tabular}{c|c|c|c}
\hline 
\textbf{Source} & \textbf{Target} & \textbf{In-distribution} & \textbf{OoD NMSE}\tabularnewline
\textbf{distribution} & \textbf{distribution} & \textbf{NMSE} & \textbf{(w/o self-ada.)}\tabularnewline
\hline 
LoS \& NLoS & LoS blockage & $-$19.99 dB & $-$19.81 dB\tabularnewline
$L=5$ paths & $L=3$ paths & $-$21.10 dB & $-$20.96 dB\tabularnewline
$L=5$ paths & $L=7$ paths & $-$18.39 dB & $-$18.22 dB\tabularnewline
\multirow{1}{*}{Hybrid-field} & \multirow{1}{*}{Near-field only} & $-$19.31 dB & $-$19.25 dB\tabularnewline
Hybrid-field & Far-field only & $-$19.43 dB & $-$19.40 dB\tabularnewline
$d_{\text{sub}}=56\lambda_{c}$ & $d_{\text{sub}}=4\lambda_{c}$ & $-$19.44 dB & $-$19.42 dB\tabularnewline
$d_{\text{sub}}=56\lambda_{c}$ & $d_{\text{sub}}=36\lambda_{c}$ & $-$19.43 dB & $-$19.42 dB\tabularnewline
$d_{\text{sub}}=56\lambda_{c}$ & $d_{\text{sub}}=76\lambda_{c}$ & $-$19.39 dB & $-$19.31 dB\tabularnewline
$d_{a}=\lambda_{c}/2$ & $d_{a}=\lambda_{c}/5$ & $-$19.94 dB & $-$19.59 dB\tabularnewline
Perfect array & Miscalibrated array & $-$19.00 dB & $-$18.90 dB\tabularnewline
\hline 
\end{tabular}
\end{table}

From the sixth to the eighth row, we show the effects of SA spacing
mismatch. In the source distribution, the SA spacing is set as $d_{\text{sub}}=56\lambda_{c}$,
while in the target distribution, it is changed to $d_{\text{sub}}=4\lambda_{c},36\lambda_{c},76\lambda_{c}$,
respectively. The change in array geometry will also affect the Rayleigh
distance, which becomes 1.44 m, 10.40 m, and 33.12 m, respectively.
The array geometry mismatch can also causes the side effect of field
mismatch. In such a complicated OoD setting, the performance of FPN-OAMP
is still robust. We further study the effect of mismatched AE spacing.
As \cite{2021Sarieddeen} pointed out, if plasmonic-based antennas
are adopted, the AE spacing can be much smaller than the conventional
choice of $\lambda_{c}/2$. In view of this, in the target distribution,
we set the AE spacing as $\lambda_{c}/5$. The OoD NMSE is very close
to the in-distribution one, suggesting that FPN-OAMP is also robust
to AE spacing mismatch. 

In the last row, we check the effects of array uncertainty. We follow
\cite{2018Ding} and focus on antenna gain miscalibration. For a perfect
array, the antenna gains are equal. For miscalibrated array, 20\%
(205) randomly picked antennas are set as $1+e_{g}$ times the normal
antenna gain with $e_{g}\sim\mathcal{N}(0,0.2)$, while the rest 80\%
(819) antennas remain unchanged. The result suggests that FPN-OAMP
is robust to array uncertainty. 

\subsection{Measurement Distribution Shifts}

\begin{figure}[t]
\centering{}\includegraphics[width=6cm]{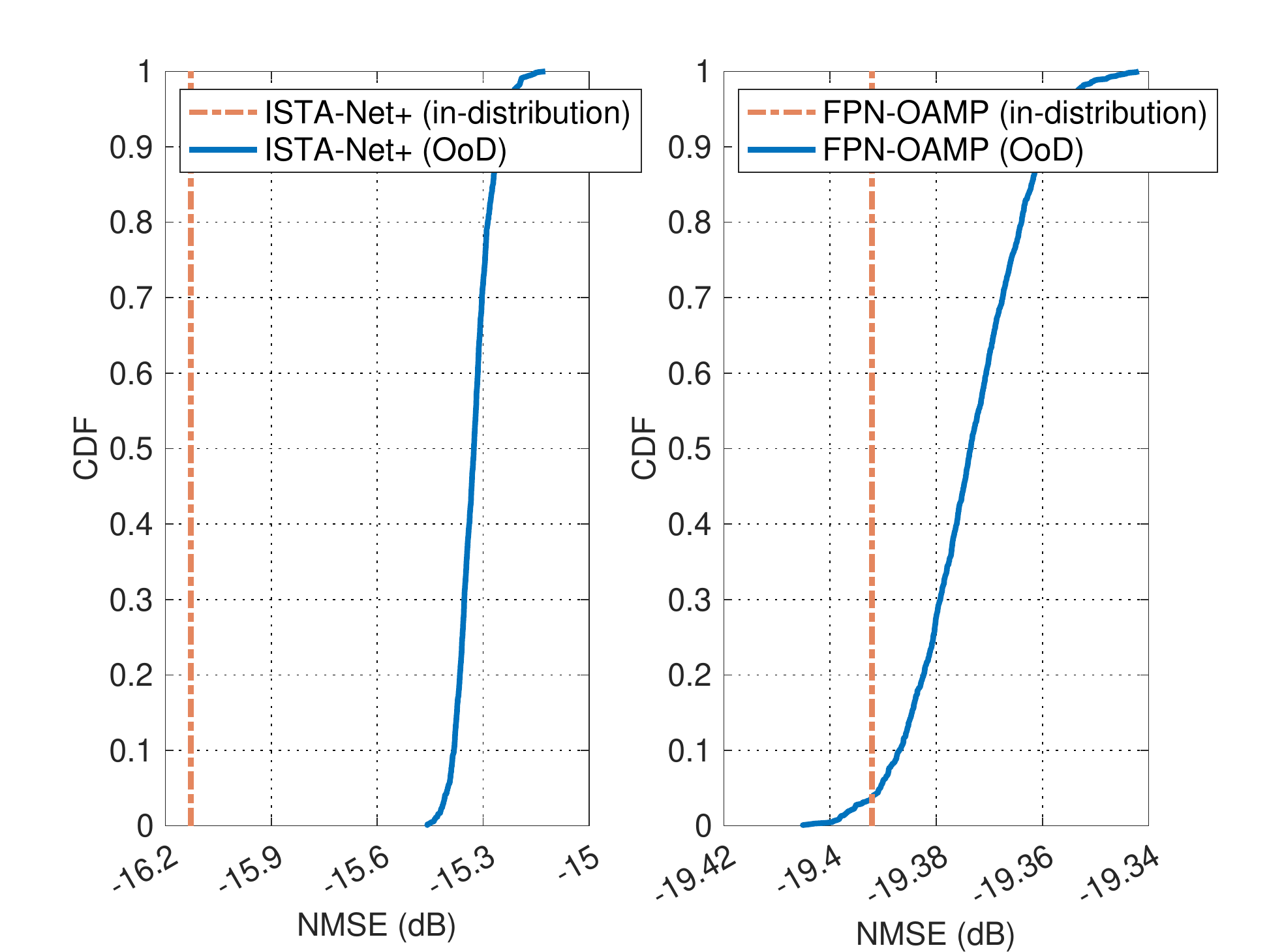}\caption{CDF of the OoD performance of ISTA-Net+ and FPN-OAMP under measurement
distribution shifts when the under-sampling ratio $\rho=50\%$. \label{fig:CDF-of-the}}
\end{figure}

In Fig. \ref{fig:CDF-of-the} and Table \ref{tab:OoD-measurement},
we present the OoD generalization and/or self-adaptation performance
of FPN-OAMP under measurement distribution shifts. The results are
all tested when the SNR is 15 dB, and the noise distribution type
is AWGN. Recall that the measurement matrix $\mathbf{M}$ is determined
by the pilot combiners $\mathbf{W}_{\text{RF},q}$, and the dictionary
$\mathbf{F}$, as in Section \ref{sec:System-Model-and}. 

Since ISTA-Net+ and FPN-OAMP are both trained using a \textit{single}
realization of the random measurement matrix, it is important to examine
whether the trained model can directly generalize to other different
realizations. In Fig. \ref{fig:CDF-of-the}, we present the cumulative
distribution function (CDF) of the performance of such kind of OoD
generalization when the under-sampling ratio $\rho$ is fixed as 50\%.
To plot the CDF, the model is tested using 1000 realizations of the
measurement matrix that are different from the one used in the training
stage. For each realization of the measurement matrix, the NMSE performance
is averaged over 5000 testing samples. As a reference, we also plot
a vertical line to show the in-distribution performance. It can be
observed that the OoD performance drop of ISTA-Net+ can be as large
as 1 dB. By contrast, the performance drop for the proposed FPN-OAMP
is almost negligible. 

In the first three rows of Table \ref{tab:OoD-measurement}, we examine
the mismatch in the under-sampling ratio $\rho=\frac{SQ}{S\bar{S}}$,
which is affected by the pilot length $Q$. In the source distribution,
the under-sampling ratio $\rho$ is set as 50\%, while in the target
distribution, it is changed to 70\%, 30\%, and 10\%, respectively.
We find that when $\rho$ is enlarged to 70\%, the OoD performance
remains unaffected. By contrast, when $\rho$ is decreased to 30\%
and 10\%, the OoD performance drops drastically. These results suggest
that FPN-OAMP can handle more received pilots but cannot directly
generalize to the case where fewer pilots are transmitted than expected.
In addition, we find that online self-adaptation can close 91\% and
98\% of the performance gap caused by the lower under-sampling ratios
for the cases of $\rho=30\%$ and $10\%$, respectively, demonstrating
its effectiveness. Note that self-adaptation is mainly intended to
handle \textit{abrupt} changes in the environment. Doing this for
every testing sample is not economical. If compatibility with different
pilot length is required, one can train FPN-OAMP with the lowest allowed
pilot length, since the model can generalize to the cases where more
pilots are transmitted. 
\begin{table}[t]
\begin{centering}
\caption{OoD generalization and self-adaptation under measurement distribution
shifts \label{tab:OoD-measurement}}
\par\end{centering}
\centering{}%
\begin{tabular}{c|c|c|c|c}
\hline 
\textbf{Source} & \textbf{Target} & \textbf{In-dist.} & \textbf{OoD NMSE} & \textbf{OoD NMSE}\tabularnewline
\textbf{dist.} & \textbf{dist.} & \textbf{NMSE} & \textbf{(w/o self-ada.)} & \textbf{(w/ self-ada.)}\tabularnewline
\hline 
$\rho=50\%$ & $\rho=70\%$ & $-$20.62 dB & $-$20.62 dB & unnecessary\tabularnewline
\multirow{1}{*}{$\rho=50\%$} & \multirow{1}{*}{$\rho=30\%$} & $-$14.86 dB & $-$7.05 dB & $-$13.16 dB\tabularnewline
$\rho=50\%$ & $\rho=10\%$ & $-$7.62 dB & $+$0.92 dB & $-$6.28 dB\tabularnewline
One-bit & Infinite-res. & $-$21.93 dB & $-$21.79 dB & unnecessary\tabularnewline
\hline 
\end{tabular}
\end{table}

In the last row of Table \ref{tab:OoD-measurement}, we further study
how the shift in the resolution of the pilot combiner affects the
performance of FPN-OAMP. In the source distribution, the elements
of the pilot combiner are picked from one-bit quantized angles, while
in the target distribution, they are instead drawn from infinite-resolution
angles. The result shows that pilot combiner resolution shift causes
almost no OoD performance drop. 

\section{Conclusions}

In this paper, we proposed FPNs, a unified and theoretically sound
framework, to design scalable, low-complexity, adaptive, and robust
DL-based channel estimation algorithms for THz UM-MIMO systems. The
unique benefits of FPNs over the prevailing deep unfolding methods
are established with firm theoretical supports. In addition to the
general framework, a specific FPN-enhanced algorithm based on OAMP,
i.e., FPN-OAMP, is also proposed. Extensive simulation results in
a typical hybrid-field THz UM-MIMO system with planar AoSA are presented
to demonstrate the significant gains of the proposed method in terms
of various key performance indicators. Furthermore, FPN-OAMP exhibits
strong robustness to distribution shifts and can directly generalize
or self-adapt to a wide range of out-of-distribution scenarios, which
makes it an ideal candidate for practical deployment. 

\appendices{}

\section{Proof of Proposition \ref{prop:implicit-gradient} \label{sec:Proof-implicit-gradient}}

The fixed point equation is $\mathbf{h}^{*}=f_{\bm{\theta}}(\mathbf{h}^{*};\mathbf{y})$,
where $\mathbf{h}^{*}$ can be viewed as an implicit function related
to $\bm{\theta}$. We denote $\mathbf{h}^{*}$ as $\mathbf{h}^{*}(\bm{\theta})$
when we treat it as an implicit function. By implicitly differentiating
both sides with respect to $\bm{\theta}$, we get
\begin{equation}
\frac{\partial\mathcal{\mathbf{h}^{*}}(\bm{\theta})}{\partial\bm{\theta}}=\frac{\partial f_{\bm{\theta}}(\mathbf{h}^{*}(\bm{\theta});\mathbf{y})}{\partial\bm{\theta}}=\frac{\partial f_{\bm{\theta}}(\mathbf{h}^{*};\mathbf{y})}{\partial\mathbf{h}^{*}}\frac{\partial\mathcal{\mathbf{h}^{*}}(\bm{\theta})}{\partial\bm{\theta}}+\frac{\partial f_{\bm{\theta}}(\mathbf{h}^{*};\mathbf{y})}{\partial\bm{\theta}}.
\end{equation}
where the second equality is due to the chain rule. Rearranging the
terms above, we reach that 
\begin{equation}
\frac{\partial\mathcal{\mathbf{h}^{*}}(\bm{\theta})}{\partial\bm{\theta}}=(\mathbf{I}-\frac{\partial f_{\bm{\theta}}(\mathbf{h}^{*};\mathbf{y})}{\partial\mathbf{h}^{*}})^{-1}\frac{\partial f_{\bm{\theta}}(\mathbf{h}^{*};\mathbf{y})}{\partial\bm{\theta}}.
\end{equation}
By using the chain rule again, we have the desired result, i.e.,
\begin{equation}
\frac{\partial\mathcal{L}}{\partial\bm{\theta}}=\frac{\partial\mathcal{L}}{\partial\mathbf{h}^{*}}\frac{\partial\mathbf{h}^{*}(\bm{\theta})}{\partial\bm{\theta}}=\frac{\partial\mathcal{L}}{\partial\mathbf{h}^{*}}(\mathbf{I}-\frac{\partial f_{\bm{\theta}}(\mathbf{h}^{*};\mathbf{y})}{\partial\mathbf{h}^{*}})^{-1}\frac{\partial f_{\bm{\theta}}(\mathbf{h}^{*};\mathbf{y})}{\partial\bm{\theta}}.
\end{equation}

\section{Proof of Theorem \ref{thm:linear-convergence}\label{sec:Proof-of-linear-convergence}}

We begin by showing that the Lipschitz constant of the LE in FPN-OAMP,
i.e., $f_{\text{LE}}(\mathbf{h}^{(t)};\mathbf{y})=(\mathbf{I}-\eta\mathbf{M}^{\dagger}\mathbf{M})\mathbf{h}^{(t)}+\eta\mathbf{M}^{\dagger}\mathbf{y}$,
equals 1. Since $f_{\text{LE}}(\mathbf{h}^{(t)};\mathbf{y})$ is an
affine mapping, its Lipschitz constant is the spectral norm, i.e.,
the largest singular value, of the matrix $(\mathbf{I}-\eta\mathbf{M}^{\dagger}\mathbf{M})$,
given by $\max_{i}\left(1-\eta\lambda_{i}\left(\mathbf{M}^{\dagger}\mathbf{M}\right)\right)$,
where $\lambda_{i}(\cdot)$ denotes the $i$-th largest eigenvalue
of a matrix. Because the non-zero eigenvalues of $\mathbf{M}^{\dagger}\mathbf{M}$
and $\mathbf{M}\mathbf{M}^{\dagger}=\mathbf{I}$ are the same, the
eigenvalues of $\mathbf{M}^{\dagger}\mathbf{M}$ equal either 0 or
1. Therefore, we can obtain that the Lipschitz constant of the LE
equals 1. According to \textbf{Lemma \ref{lem:The-composition-of}},
the Lipschitz constant of $f_{\bm{\theta}}(\cdot;\mathbf{y})=(f_{\text{NLE},\bm{\theta}}\circ f_{\text{LE}})(\cdot;\mathbf{y})$
is the same as that of $f_{\text{NLE},\bm{\theta}}(\cdot)$, i.e.,
$L$. Further applying \textbf{Lemma \ref{lem:Banach}} yields the
desired result. 

\bibliographystyle{IEEEtran}
\bibliography{references_THz_TWC}

\end{document}